\documentclass[12pt]{article}

\usepackage[table]{xcolor}

\usepackage{times}
\usepackage{indentfirst}
\usepackage{setspace}
\usepackage{natbib}
\usepackage[pdftex]{graphicx}
\usepackage{subfigure}
\usepackage{makecell}
\usepackage{booktabs}
\usepackage{array}
\usepackage{url}
\usepackage{algorithm}
\usepackage{algorithmic}
\usepackage{bm}
\usepackage{smile}
\usepackage{mathtools}
\usepackage{wrapfig}
\usepackage{lipsum}
\usepackage{mathrsfs}
\usepackage{rotating} 
\usepackage{float}
\usepackage{multirow}
\usepackage{apalike}
\usepackage{blindtext}

\definecolor{grey}{rgb}{0.5,0.5,0.5}

\def\reals{{\mathbb R}}

\def\E{{\mathbb E}}
\def\argmin{\mathop{\text{\rm arg\,min}}}
\def\argmax{\mathop{\text{\rm arg\,max}}}

\def\span{\text{span}}

\makeatletter
\newcommand*{\addFileDependency}[1]{%
  \typeout{(#1)}
  \@addtofilelist{#1}
  \IfFileExists{#1}{}{\typeout{No file #1.}}
}
\makeatother

\newcommand{\blind}{0}

\begin{document}

\def\spacingset#1{\renewcommand{\baselinestretch}%
{#1}\small\normalsize} \spacingset{1}


\if0\blind
{
  \title{\bf Few-Round Distributed Principal Component Analysis: Closing the Statistical Efficiency Gap by Consensus}
  \author{ZeYu Li\\
    Department of Statistics, Fudan University, China\\
    Xinsheng Zhang \\
    Department of Statistics, Fudan University, China\\
    and \\
    Wang Zhou\\
    Department of Statistics and Data Science, \\
    National University of Singapore, Singapore}
  \maketitle
} \fi

\if1\blind
{
  \bigskip
  \bigskip
  \bigskip
  \begin{center}
    {\LARGE\bf Few-Round Distributed Principal Component Analysis: Closing the Statistical Efficiency Gap by Consensus}
\end{center}
  \medskip
} \fi

\begin{abstract}
Distributed algorithms and theories are called for in this era of big data. Under weaker local signal-to-noise ratios, we improve upon the celebrated one-round distributed principal component analysis (PCA) algorithm designed in the spirit of divide-and-conquer, by introducing a few additional communication rounds of consensus. The proposed shifted subspace iteration algorithm is able to close the local phase transition gap, reduce the asymptotic variance, and also alleviate the potential bias. Our estimation procedure is easy to implement and tuning-free. The resulting estimator is shown to be statistically efficient after an acceptable number of iterations. We also discuss extensions to distributed elliptical PCA for heavy-tailed data. Empirical experiments on synthetic and benchmark datasets demonstrate our method's statistical advantage over the divide-and-conquer approach.
\end{abstract}

\noindent%
\bigskip
\spacingset{1.8} 

\section{Introduction}

Principal component analysis (PCA) remains the cornerstone technique for dimensionality reduction \citep{pearson1901liii}, with applications in regression, functional analysis, and factor modeling \citep{jolliffe2003modified, fan2013large, chen2021statistical, he2022large}. The primary objective of principal component analysis is to lower the dimensionality of a dataset while preserving the maximal amount of original variance. This is achieved by computing the eigenvectors and eigenvalues of the sample covariance matrix, where the eigenvectors correspond to the principal directions of variance, and the eigenvalues quantify the extent of variability along these directions.

In this era of massive information, large datasets are often scattered across distant machines. The attempt to fuse or aggregate these datasets is extremely difficult. Over the past two decades, significant progress has been made in the development of distributed PCA, contributing to the broader efforts to create accurate and efficient distributed statistical methods \citep{wu2018review}. In general, there are two data partition regimes for distributed PCA. For horizontal partition, each local machine contains all the features of a subset of subjects \citep{qu2002principal}. Meanwhile, for vertical partition, each machine has a subset of features of the same group of subjects. It has roots in sensor networks and signal processing \citep{kargupta2001distributed,schizas2015distributed}. This work focuses on horizontal partition. However, our algorithm could also be applied to the vertical partition settings, thanks to the symmetry of columns and rows in rank $r$ approximation \citep{qu2002principal,fan2019distributed}.

For horizontal partition, \cite{fan2019distributed} proposed a novel one-round distributed principal component analysis algorithm. In the one-round algorithm, the local machines only send the leading eigenvectors to the central machine, on which the local subspace estimators are integrated. Following \cite{fan2019distributed}, \cite{xu2022distributed} propose a distributed sufficient dimension reduction algorithm using the eigenspace integration. Meanwhile, in \cite{he2022distributed}, the authors discuss the distributed PCA problem under potentially heavy-tailed data, and propose to calculate sample sample Kendall'$\tau$ matrices locally \citep{Fan2018LARGE,han2018eca}. 

The methods in \cite{fan2019distributed,xu2022distributed,he2022distributed} are all designed in the spirit of divide and conquer for subspace estimators, where the central machine aggregates the locally estimated subspaces using a notion of ``subspace center'' in a least square sense. However, for divide-and-conquer methods, over-splitting of datasets might cause problems: conceptually, if the local sample size is too small, then the local estimators are less reliable and eventually result in a larger bias in the aggregated estimator. Due to this reason, the number of splitted machines is often restricted in these works, so that the bias term is of a higher order and can be omitted.

Subsequently, \cite{chen2022distributed} proposed a multi-round algorithm that enjoys a linear convergence rate under weaker restrictions and is further applied to principal component regression and single index model. Their approach has no restriction on the number
of machines, as it is eventually unbiased. In \cite{chen2022distributed}, the authors avoid transmitting the Hessian matrices in the iterations by only using the Hessian information on the first machine, which is a natural idea \citep{shamir2014communication,jordan2019communication,fan2023communication}, but their method involves both inner and outer iterations, resulting in higher communication costs; see also \cite{chen2024distributed} for distributed estimation of canonical correlation analysis that is essentially unbiased as well. An alternative multi-round method is given by \cite{garber2017communication}, which mainly estimates the first eigenvector. For other related works, one might see \cite{charisopoulos2021communication,huang2021communication} and the references therein.

\subsection{Problem Formulation}\label{sec:setting}

Going down the stream of the innovative one-round procedure \citep{fan2019distributed}, our target is to solve the distributed PCA problem in a statistically efficient manner, with as small number of iteration rounds as possible.

In fact, under relatively high local signal-to-noise ratio (SNR) conditions, the one-round estimator exhibits matching statistical error rates with the pooling PCA, i.e., PCA over the full sample \citep{fan2019distributed}, as the bias term is of a higher order. By local signal-to-noise ratio, we describe the amount of statistical information stored in each local machine. If the local signal-to-noise ratio tends to infinity, then each local machine contains enough information to recover the principal subspace alone, and distributed learning might not be necessary. On the other hand, when the local datasets have weaker signal-to-noise ratios, due to small sample sizes or weak signals, existing one-round subspace integration theory no longer holds, as it requires the local quadratic expansion of the subspace estimation error, i.e., the higher-order Davis-Kahan theorem \citep{fan2019distributed,li2024tpca}. 

However, it is believed that the weak or moderate local SNR could be found in many real-world problems, as datasets are often scattered across a large number of local machines such as smartphones and personal computers, each containing a relatively small sample size. In these cases, the one-round estimator might not be statistically efficient, due to the local phase transition gap, the inflated variance, and the inherent bias as revealed later in this work.

This article examines method performance under weaker local signal-to-noise ratio conditions. We follow \cite{fan2019distributed} and assume homogeneous sub-sample sizes. Namely, we have $K$ machines, each containing $n$ observations, denoted as $\{x_{i}^{k}\}^{k\in[K]}_{i\in[n]}$. We assume that $x_{i}^{k}\in\reals^p$ are i.i.d. random vectors with zero mean and a homogeneous population covariance matrix $\Sigma$, whose leading $r$ eigenvectors form a $p\times r$ column orthogonal matrix $U:=(u_1,\dots,u_r)$. The non-increasing eigenvalues of $\Sigma$ are denoted as $\lambda_i$, for $i\in [p]$. We assume that $\lambda_r>\lambda_{r+1}$. Principal component analysis aims to retrieve $U$ from noisy observations. This work focuses on the weaker local signal-to-noise ratio regime where $n$ might be of the same order as $p$, while $r$ and $\{\lambda_i\}_{i=1}^{r}$ are treated as constants. In this regime, random matrix theory provides accurate analytical tools \citep{marchenko1967distribution,bai2010spectral,couillet2022random}.

\subsection{Main Contributions and Notations}

This work is an extension of \cite{fan2019distributed} towards the previously unexplored regime of weaker local SNR. Random matrix theory reveals the existence of a local phase transition gap, below which the pooling PCA estimator can be consistent as $K\rightarrow\infty$, but the divide-and-conquer estimator is questionable. Even above the local phase transition gap, the one-round estimator still exhibits higher asymptotic subspace variance than the pooling estimator, not to mention the potential bias that can be of constant order. The result concerning the inflated subspace variance of the divide-and-conquer estimator above the local phase transition threshold is, to the best of our knowledge, new in the literature.

To close the phase transition gap, reduce the variance and alleviate the potential bias of the one-round estimator, we suggest a few additional rounds of consensus, via distributed subspace iteration with eigenvalue shift. Different from the divide-and-conquer estimators where only local subspace information is sent to the central machine, in further consensus rounds, the local machines encode shared information by receiving the aggregated subspace estimator from the previous round, and send more reliable information to the central machine. Compared to the existing multi-round algorithms \citep{garber2017communication,chen2022distributed}, our method does not require inner and outer iterations; it is easy to implement and tuning-free. Furthermore, the resulting estimator is shown to be statistically efficient after an acceptable number of iterations.

The theoretical performance of estimators from the first divide-and-conquer round and further consensus rounds is discussed and compared under different models, including the popular Gaussian spiked model \citep{paul2007asymptotics}, more general cases where the one-round estimator is biased \citep{fan2019distributed}, and also the elliptical spiked model which includes heavy-tailed distributions and finds applications in finance and macroeconomics \citep{Fan2018LARGE,han2018eca,he2022distributed}.

The proposed method is then applied on synthetic datasets and benchmark tabular datasets. Besides validating our theoretical arguments, we show that our algorithm has a persistent statistical advantage over the one-round algorithm in the empirical experiments. The local phase transition gap is also observed in two of the ten benchmark datasets, i.e., MiniBooNE and road\_safety.

To end this section, we introduce some notations. For a real symmetric matrix $A$, we write $\|A\|_{2}$ as the operator norm of the matrix $A$, and write $\|A\|_{F}$ as its Frobenius norm. Meanwhile, $\|a\|$ represents the $\ell_2$ norm of a vector $a$. The $o_{p}$ is for convergence to zero in probability and the $O_{p}$ is for stochastic boundedness. For any $a\in \reals$, $\lceil a \rceil$ means the smallest integer that is not smaller than $a$. We write $x\lesssim y$ if $x\leq Cy$ for some $C>0$, $x\gtrsim y$ if $x\geq cy$ for some $c>0$, and $x\asymp y$ if both $x\lesssim y$ and $x\gtrsim y$ hold.  Note that the constants $c$ and $C$ may not be identical in different lines.

\section{Methodology}

Under the settings in Section \ref{sec:setting}, if we transfer all data samples (if $p\geq n$), or local sample covariance matrices (if $p<n$), to the central machine, we obtain the following pooled sample covariance matrix
$$\hat{\Sigma}^{\cP}:=\frac{1}{N}\sum_{k=1}^{K}\sum_{i=1}^{n}x_i^k(x_i^k)^{\top},$$ 
with the communication cost of $O[Kp\min(p,n)]$. Its leading $r$ eigenvectors form a $p\times r$ column orthogonal matrix $\hat{U}^{\cP}:=(\hat{u}_1^{\cP},\dots,\hat{u}_r^{\cP})$. We call $\hat{U}^{\cP}$ the pooling PCA estimator. It stands as the goal of the distributed PCA algorithms. 

According to \cite{fan2019distributed}, we do not need to transfer all data to the central machine, but can respectively calculate local sample covariance matrices $\tilde{\Sigma}_k:=\sum_{i=1}^{n}x_i^k(x_i^k)^{\top}/n$, for $k\in[K]$
whose leading $r$ eigenvectors form a $p\times r$ column orthogonal matrix $\tilde{U}_k:=(\tilde{u}^k_1,\dots,\tilde{u}^k_r)$. We only send $\tilde{U}_k$ to the central machine, resulting in the total communication cost of $O(Kpr)$. Then, the central machine calculates the leading $r$ eigenvectors of
$$\hat{\Sigma}^{(1)}:=\frac{1}{K}\sum_{k=1}^{K}\tilde{U}_k\tilde{U}^{\top}_k,$$
denoted as $\hat{U}^{(1)}:=(\hat{u}_1^{(1)},\dots,\hat{u}_r^{(1)})$, which is called the one-round distributed PCA estimator in this article. Here the superscript $(1)$ means one round of communication. In fact, $\hat{U}^{(1)}$ is the solution of the following optimization problem:
\begin{equation*}
\begin{aligned}
         \hat{U}^{(1)}(\hat{U}^{(1)})^{\top}&=\argmax_{UU^{\top}\in \cG(p,r)}\frac{1}{K} \sum_{k=1}^{K} \tr(\tilde{U}_k\tilde{U}_k^{\top}UU^{\top}) \\
        &= \argmin_{UU^{\top}\in \cG(p,r)}\frac{1}{K} \sum_{k=1}^{K}\left\|\tilde{U}_k\tilde{U}_k^{\top}-UU^{\top}\right\|^2_F,
\end{aligned}
\end{equation*}
where $\mathcal{G}(p,r)$ is defined as the set of rank-$r$ linear projectors of $\mathbb{R}^p$. Note that any subspace can be uniquely represented by the orthogonal projector from $\mathbb{R}^p$ onto itself. Hence, $\hat{U}^{(1)}(\hat{U}^{(1)})^{\top}$ can be viewed as the subspace center, or average, of the local subspace estimators $\{\tilde{U}_k\tilde{U}_k^{\top}\}_{k=1}^{K}$, in the sense of the projection metric.

Owing to its geometric clarity, similar approaches have been separately introduced by multiple researchers in diverse disciplines. For example, \cite{crone1995statistical} apply the concept of a subspace center in the context of satellite meteorology. Similarly, \cite{liski2016combining} adopt this strategy to integrate results from several dimensionality reduction techniques. In the end, the subspace center can also be interpreted as a specific case of the extrinsic Fr\'{e}chet mean \citep{bhattacharya2003large}.

\subsection{Shifted Subspace Iteration}
In this work, we suggest adding a few shifted subspace iteration rounds after acquiring the one-round estimator. After acquiring $\hat{U}^{(1)}$, the central machine sends $\hat{U}^{(1)}$ back to the local machines, so that the local machines can share the same global knowledge. Then, for $\hat{U}_{\perp}^{(1)}(\hat{U}_{\perp}^{(1)})^{\top}=I_p-\hat{U}^{(1)}(\hat{U}^{(1)})^{\top}$, the $k$-th local machine calculates
$$\tilde{G}^{(1)}_k:= \tilde{\Sigma}_k \hat{U}^{(1)}-(\tilde{\sigma}^{(1)}_{k})^{2} \hat{U}^{(1)} \in \reals^{p\times r},\quad \text{where}\quad (\tilde{\sigma}^{(1)}_{k})^{2}:=\frac{1}{p-r}\tr\left(\tilde{\Sigma}_k\hat{U}_{\perp}^{(1)}(\hat{U}_{\perp}^{(1)})^{\top}\right),$$
and sends $\tilde{G}^{(1)}_k$ to the central machine. The central machine orthogonalizes $\sum_{k=1}^{K}\tilde{G}^{(1)}_k/K$ using QR factorization, so as to acquire $\hat{U}^{(2)}$. Clearly, this procedure can be further iterated for any given iteration rounds $T$, with the communication cost of $O(TKpr)$; see Algorithm \ref{alg:frdpca} for details. Before we rigorously discuss the theoretical properties of the few-round estimator $\hat{U}^{(t)}$, some intuitions are stated as follows.

\begin{algorithm}
\caption{Few-round distributed principal component analysis.}\label{alg:frdpca}
\begin{algorithmic}[1]
\REQUIRE ~~\\
    $\tilde{\Sigma}_k$ for $k\in[K]$; cut-off dimension $r$; iteration rounds $T\geq 2$;\\
\ENSURE ~~\\
    \STATE the first round: the local machines send the local leading eigenvectors $\tilde{U}_k$ to the central machine; the central machine acquires the one-round distributed PCA estimator $\hat{U}^{(1)}$ by calculating the leading $r$ eigenvectors of $\hat{\Sigma}^{(1)}=\sum_{k=1}^{K}\tilde{U}_k\tilde{U}^{\top}_k/K$;
    \STATE shifted subspace iteration: given $\hat{U}^{(t)}$ for $t\geq 1$, the central machine sends $\hat{U}^{(t)}$ back to the local machines; the local machines send $\tilde{G}^{(t)}_k:=\tilde{\Sigma}_k \hat{U}^{(t)}-(\tilde{\sigma}^{(t)}_{k})^{2} \hat{U}^{(t)}$ to the central machine, where $(\tilde{\sigma}^{(t)}_{k})^{2}:=\tr(\tilde{\Sigma}_k\hat{U}_{\perp}^{(t)}(\hat{U}_{\perp}^{(t)})^{\top})/(p-r)$; the central machine obtains $\hat{U}^{(t+1)}$ by orthogonalizing $\sum_{k=1}^{K}\tilde{G}^{(t)}_k/K$ using QR factorization.
\RETURN Few-round distributed PCA estimator $\hat{U}^{(T)}$.
\end{algorithmic}
\end{algorithm}

 First, note that in the one-round algorithm, the local machines send $\tilde{U}_k$, which is calculated using the $k$-th local dataset only. Meanwhile, in further iterations of Algorithm \ref{alg:frdpca}, the local machines are aware of the information from other machines via $\hat{U}^{(t-1)}$ sent back from the central machine, and $\tilde{G}^{(t-1)}_k$ encodes shared information by consensus. This slight difference of consensus closes the phase transition gap discussed later in Table \ref{tab:method_comparison} and Figure \ref{fig-phase}, and also enhances the statistical accuracy above the local phase transition threshold by reducing the variance and alleviating the bias, achieving efficiency.

Second, we can see that in the $(t+1)$-th round, the central machine eventually gets
\begin{equation}\label{eq:2rd-pool}
  \left(\sum_{k=1}^{K}\tilde{\Sigma}_k/K-\frac{\sum_{i=1}^{K}(\tilde{\sigma}^{(t)}_{k})^{2}}{K}I_p\right) \hat{U}^{(t)}=\left(\hat{\Sigma}^{\cP}-(\hat{\sigma}^{(t)})^{2}I_p\right)\hat{U}^{(t)}, \quad\text{where}
\end{equation}
$$(\hat{\sigma}^{(t)})^{2}:=\sum_{i=1}^{K}(\tilde{\sigma}^{(t)}_{k})^{2}/K=\frac{1}{K(p-r)}\sum_{k=1}^{K}\tr\left(\tilde{\Sigma}_k\hat{U}_{\perp}^{(t)}(\hat{U}_{\perp}^{(t)})^{\top}\right)=\frac{1}{p-r}\tr\left(\hat{\Sigma}^{\cP}\hat{U}_{\perp}^{(t)}(\hat{U}_{\perp}^{(t)})^{\top}\right).$$
Namely, the central machine almost acquires the information of the pooled sample covariance matrix $\hat{\Sigma}^{\cP}$ painlessly. In fact, \eqref{eq:2rd-pool} resembles the subspace iteration when calculating the eigenvectors in numerical algebra \citep{ford2014numerical}. It is foreseeable that if the iteration proceeds, $\hat{U}^{(t)}$ shall tend to the leading eigenvalues of $\hat{\Sigma}^{\cP}$ as $t\rightarrow \infty$, at a linear convergence rate. 

Third, the numerical convergence rate of subspace iteration is closely related to the eigenvalue ratio of the target matrix of interest. For principal component analysis, the spiked eigenvalues are often much larger than the noise eigenvalues, so subspace iteration converges fast. However, in our particular case of weaker local signal-to-noise ratio, the signal and noise eigenvalues are of the same order. Hence, it is often beneficial to shift the noise eigenvalues towards $0$, in order to accelerate the convergence. Faster numerical convergence is important in the context of distributed learning, as each iteration step requires an additional round of communication and hence more costs.

For illustration, one may consider the spiked model with uniform noise eigenvalues \citep{paul2007asymptotics,baik2006eigenvalues,bai2018consistency}, i.e., $\lambda_{r+1}=\dots=\lambda_p=1$. Without eigenvalue shift, the eigenvalue ratio of subspace iteration will be $|1/\lambda_{r}|>0$. Meanwhile, if we carefully shift the noise eigenvalues to $0$, the eigenvalue ratio also tends to $0$ and faster convergence is guaranteed. In fact, even for non-uniform noise eigenvalues \citep{wang2017asymptotics,cai2020limiting}, the noise jointly affects the signals by the quantity $\sigma^2 = \sum_{j=r+1}^{p}\lambda_j/(p-r)$ under divergent spikes, which is exactly the term subtracted by Algorithm \ref{alg:frdpca} in order to speed up convergence; one can refer to, e.g., Theorem 3.1 of \cite{wang2017asymptotics} for details.

Finally, the choice of iteration times $T$ is essentially a trade-off between the statistical error and communication costs, and it depends on each specific problem at hand. As shown by our experiments on the benchmark datasets, after two to three rounds of communication, further iterations typically have a marginal effect on the statistical accuracy. Under communicational constraints, $T\in\{3,4,5\}$ could be a set of ``rule of thumb'' choices.

\section{Theory}

In this section, we discuss the statistical advantages gained from the additional consensus rounds after acquiring the one-round divide-and-conquer estimator $\hat{U}^{(1)}$. It is shown that further iterations can close the local phase transition gap, reduce the asymptotic subspace variance, and also alleviate the potential bias of $\hat{U}^{(1)}$, under weaker local SNR. 

We begin by introducing the notion of asymptotic efficiency in the context of distributed PCA. In particular, we say any distributed PCA estimator $\hat{U}$ is asymptotically efficient if and only if it is as good as the pooling PCA estimator in an asymptotic subspace variance manner.

\begin{definition}[Asymptotic efficiency]
    For any distributed PCA estimator $\hat{U}$ and the pooling PCA estimator $\hat{U}^{\cP}$, we define the asymptotic subspace variance ratio as  $$\mathcal{V}\left(\hat{U}\right):=\frac{\E\left(\left\|\hat{U}\hat{U}^{\top}-UU^{\top}\right\|^2_F\right)}{\E\left(\left\|\hat{U}^{\cP}(\hat{U}^{\cP})^{\top}-UU^{\top}\right\|^2_F\right)}.$$
    We say the distributed PCA estimator $\hat{U}$ is asymptotically efficient if and only if
    $$\lim\limits_{k\rightarrow \infty} \lim\limits_{n,p\rightarrow \infty} \mathcal{V}\left(\hat{U}\right) = 1.$$
\end{definition}

In Section \ref{sec:var}, we first work under the classical Gaussian spiked model \citep{paul2007asymptotics} as given in the following Assumption \ref{assump:1}, that $\hat{U}^{(1)}$ is unbiased. This model provides us with a clear picture of what may happen as the spike strength $l_r$ gradually grows from $0$ to infinity, given a fixed $c=p/n>0$. In Table \ref{tab:method_comparison}, we collect the theoretical results corresponding to different signal-to-noise regimes of the Gaussian spiked model. Here is a brief outline of the results in Table \ref{tab:method_comparison}, which are also numerically verified in Figure \ref{fig-phase} in the Appendix.

\begin{table}
\centering
\caption{Theoretical comparisons of different methods under various signal-to-noise regimes of the Gaussian spiked model, given a fixed $c=p/n>0$.}
\label{tab:method_comparison}
\resizebox{0.9\textwidth}{!}{%
\begin{tabular}{|c|c|c|c|}
\hline
Methods & Pooling PCA & One-round DPCA & Few-round DPCA \\ \hline
\cellcolor{grey!30}{$l_r\in(0,\sqrt{c/K}]$} & \cellcolor{grey!30}{Below Pool Threshold} & \cellcolor{grey!30}{Below Pool Threshold} & \cellcolor{grey!30}{Below Pool Threshold} \\ \hline
\cellcolor{yellow!30}{$l_r\in(\sqrt{c/K},\sqrt{c}]$} & \cellcolor{yellow!30}{Efficient} &  \cellcolor{yellow!30}{Below Local Threshold} & \cellcolor{yellow!30}{Efficient if $T\gtrsim\log(K)$}\\ \hline
\cellcolor{green!30}{$l_r>\sqrt{c}$} &\cellcolor{green!30}{Efficient}  & \cellcolor{green!30}{Not Efficient} & \cellcolor{green!30}{Efficient if $T=2$}   \\ \hline
\cellcolor{pink!30}{$l_r\gg \sqrt{c}$} &\cellcolor{pink!30}{Efficient }&\cellcolor{pink!30}{Efficient} &\cellcolor{pink!30}{Efficient}  \\ \hline
\end{tabular}
}
\end{table}

\begin{itemize}
    \item In the gray regime (of both Table \ref{tab:method_comparison} and Figure \ref{fig-phase} below) where $l_r\in(0,\sqrt{c/K}]$, the signal strength is beneath the phase transition threshold of pooling PCA as depicted by Proposition \ref{prop:rmt-eigenvector}, and no method is able to retrieve the signal eigenvector, unless additional model structures are imposed.
    \item For $l_r\in(\sqrt{c/K},\sqrt{c}]$, i.e., the yellow regime, the divide-and-conquer estimator loses eigenvector information due to the higher local phase transition threshold after the over-splitting of data, as claimed by Corollary \ref{coro:lpt}. Meanwhile, the few-round estimator is able to retrieve some of the subspace information after reasonable rounds of iterations, even if the initialization is almost random, as supported by Theorem \ref{theorem:3}. This phase transition gap is also observed in some of the real data cases, as shown later in Figure \ref{fig:cases}.
    \item The green regime where $l_r>\sqrt{c}$ corresponds to the settings of both Theorem \ref{theorem:1rdpca} and Theorem \ref{theorem:2rdpca}, i.e., relatively weaker local signal-to-noise ratio above the local phase transition threshold. We show that the one-round estimator is consistent as $K\rightarrow \infty$, but still not asymptotically efficient (has inflated variance). Fortunately, merely an additional round of consensus suffices to close this efficiency gap.

    \item Finally, in the pink regime such that $l_r\gg \sqrt{c}$, the local machines tends to consistently estimate the signal eigenvector, and all methods are efficient. Also, a large number of the benchmark tabular datasets fall in this regime.
\end{itemize}

In Section \ref{sec:var}, we also discuss the by-products of Algorithm \ref{alg:frdpca}, including distributed spiked eigenvalue estimation and the asymptotic distribution of the bilinear forms, whose asymptotic variance can also be estimated in a distributed manner. Then in Section \ref{sec:bias}, we work on more general cases in which the one-round estimator is generally biased. We show that the proposed few-round algorithm is able to alleviate the bias, which is often the main term under weaker local SNR above the local phase transition threshold. Finally, in Section \ref{sec:epca}, we discuss the extension to distributed robust elliptical PCA, which can find applications in finance and macroeconomics \citep{Fan2018LARGE,han2018eca,he2022distributed}.

\subsection{Gaussian Spiked Model}\label{sec:var}
In this section, we first discuss under the classical Gaussian spiked model \citep{paul2007asymptotics}, that $\hat{U}^{(1)}$ is unbiased.


\begin{assumption}[Gaussian spiked model \citep{paul2007asymptotics}] \label{assump:1}
Assume that $x_{i}^{k}\in\reals^p$ are i.i.d. Gaussian random vectors with zero mean and a homogeneous population covariance matrix $\Sigma= \sum_{i=1}^{r} l_i u_iu_i^{\top} + I_p$. Assume that $p/n\rightarrow c$ for some $c>0$ as $n$, $p\rightarrow \infty$, while $l_1>l_2>\dots>l_r>0$.
\end{assumption}

It is possible to generalize Assumption \ref{assump:1} above. For instance, we can relax the Gaussian assumption to, e.g., finite fourth moments, together with conditions like symmetric innovation (it ensures the unbiasedness of $\hat{U}^{(1)}$) as in \cite{fan2019distributed}, keeping almost all arguments in this section unchanged. However, one of the main purposes of this section is to show that $\hat{U}^{(1)}$ via divide-and-conquer might lose statistical efficiency under weaker local SNR due to, e.g., over-splitting of the dataset. Meanwhile, the proposed algorithm can close this efficiency gap by rapidly approaching $\hat{U}^{\cP}$ by consensus. Assumption \ref{assump:1} suffices for such a purpose and saves us from heavy notations, providing a clearer picture.

Due to the celebrated phase transition phenomenon in random matrix theory \citep{paul2007asymptotics,couillet2022random}, the $r$-th eigenvector information is lost for the local machines unless the spike $l_r$ exceeds the local phase transition threshold, namely $l_r > \sqrt{c}$. We restate this well-known result in the following Corollary \ref{coro:lpt} for the sake of convenience. As a result, if the signal strength is below this threshold, then the performance of the one-round distributed PCA estimator cannot be ensured, even if the number of machines $K\rightarrow \infty$.


\begin{corollary}[Local phase transition]\label{coro:lpt}
Under Assumption \ref{assump:1}, for the $k$-th local dataset, let $\tilde{U}_k:=(\tilde{u}^k_1,\dots,\tilde{u}^k_r)$ be the leading $r$ eigenvectors of the local sample covariance matrices $\tilde{\Sigma}_k$. If $l_r\leq\sqrt{c}$, we have $u_r^{\top}\tilde{u}^k_r  \xrightarrow{\text {a.s.}} 0$.
\end{corollary}

Meanwhile, for pooling PCA to work, one only needs $l_r> \sqrt{c/K}$. Clearly, the local phase transition threshold is higher than the phase transition threshold of pooling PCA, so there is also a theoretical phase transition gap, i.e., $\sqrt{c/K}<l_r\leq \sqrt{c}$, in which pooling PCA can be consistent as $K\rightarrow\infty$, but the one-round distributed PCA is not. That is, if we over-split the dataset, the local SNR can be so small that the local subspace estimators are no longer reliable, which is troublesome for divide-and-conquer methods. The phase transition gap is observed in both synthetic datasets (Figure \ref{fig-phase}) and some real data cases (Figure \ref{fig:cases}).

Then, we focus on the cases above the local phase transition threshold, i.e., $l_r>\sqrt{c}$. In this case, the one-round distributed PCA is consistent as $K\rightarrow \infty$. However, it is shown in Theorem \ref{theorem:1rdpca} that it still has an inflated asymptotic variance, hence not statistically efficient unless $l_r\gg \sqrt{c}$, i.e., each local machine contains sufficient information to consistently retrieve the population subspace of interest.

\begin{theorem}[One-round distributed PCA]\label{theorem:1rdpca}
    Under Assumption \ref{assump:1}, if $l_r>\sqrt{c}$, we have
    $$\lim\limits_{k\rightarrow \infty} \lim\limits_{n,p\rightarrow \infty} \mathcal{V}\left(\hat{U}^{(1)}\right) = \frac{\sum_{i=1}^{r}(pl_i^{-1}+pl_i^{-2})/(n-pl_i^{-2})}{\sum_{i=1}^{r}  (pl_i^{-1}+p l_i^{-2})/n}>1.$$
\end{theorem}

From Theorem \ref{theorem:1rdpca}, we notice that the asymptotic subspace variance ratio of the one-round distributed PCA estimator depends heavily on the local SNR, which is characterized by $n$, the local sample size, $p$, the data dimension, and $l_i$, the signal strength. If $n\gg pl_r^{-2}$, i.e., $l_r\gg \sqrt{c}$ and the local signal-to-noise ratio diverges, then we have $$\lim_{k\rightarrow \infty} \lim_{n,p\rightarrow \infty} \mathcal{V}\left(\hat{U}^{(1)}\right)\rightarrow 1,$$
and the one-round PCA estimator tends to be asymptotically efficient. On the other side, due to the phase transition phenomenon in random matrix theory, if $n\leq pl_r^{-2}$, i.e., $l_r\leq \sqrt{c}$, the $r$-th eigenvector information is lost locally, as shown by Corollary \ref{coro:lpt} above. In Theorem \ref{theorem:1rdpca}, as $n\rightarrow pl_r^{-2}$ from above, i.e., $l_r\rightarrow \sqrt{c}$ and approaches the phase-transition threshold, we shall have 
$$\lim_{k\rightarrow \infty} \lim_{n,p\rightarrow \infty} \mathcal{V}\left(\hat{U}^{(1)}\right)\rightarrow \infty.$$
In conclusion, the one-round distributed PCA estimator has a larger asymptotic statistical error than the pooling PCA estimator, unless the local signal-to-noise ratio diverges.

Fortunately, we show as follows that the second round of consensus suffices to close the statistical efficiency gap, by significantly reducing the asymptotic subspace variance. Indeed, $\hat{U}^{(2)}$ from Algorithm \ref{alg:frdpca} is asymptotically efficient under Assumption \ref{assump:1} above the local phase transition threshold.

\begin{theorem}[Variance reduction]\label{theorem:2rdpca}
    Under Assumption \ref{assump:1}, if $l_r>\sqrt{c}$, we have
    $$\lim\limits_{k\rightarrow \infty} \lim\limits_{n,p\rightarrow \infty} \mathcal{V}\left(\hat{U}^{(2)}\right) =1.$$
\end{theorem}

We point out that another by-product of the two-round distributed PCA is the spiked eigenvalue estimation. In one-round distributed PCA, the local eigenvalue information is discarded. In the second round, on the other hand, the central machine is able to receive the signal eigenvalue information from the local machines.

\begin{corollary}[Spiked eigenvalue estimation]\label{coro:eigenvalues}
     Under Assumption \ref{assump:1}, for $i\in [r]$, letting $\hat{l}^{(2)}_i$ be the $i$-th largest singular value of $\sum_{k=1}^{K}\tilde{G}^{(1)}_k/K$, if $l_r>\sqrt{c}$, we have $\hat{l}_i^{(2)}-l_i = O_p(K^{-1/2})$.
\end{corollary}

In \cite{koltchinskii2016asymptotics}, the authors analyze the asymptotics of bilinear forms of empirical spectral projectors for Gaussian data. Given the results above, we are able to discuss the same problem in the distributed learning context; see also \cite{bao2022statistical} for a detailed study of the asymptotic behavior of extreme eigenvectors under general conditions. For $u$, $v\in S^{p-1}$, the $(p-1)$-dimensional sphere, we are interested in the asymptotic distribution of the bilinear form $\langle u, [\hat{U}^{(t)}(\hat{U}^{(t)})^{\top}-UU^{\top}]v\rangle$. The result is stated in Corollary \ref{coro:an}.

\begin{corollary}[Asymptotic normality of the bilinear form]\label{coro:an}
Under Assumption \ref{assump:1}, for $u$, $v\in S^{p-1}$, given $\sigma_{(u,v)}^2$ and $(\hat{\sigma}_{(u,v)}^{(t)})^2$ defined later in \eqref{eq:sigma2} and \eqref{eq:hatsigma2}, if $l_r>\sqrt{c}$ and $\sigma_{(u,v)}^2>0$, we have
$$ \hat{U}^{(2)}(\hat{U}^{(2)})^{\top}-UU^{\top}=\hat{L}^{(2)} + \hat{R}^{(2)},\quad \text{where}$$
$$\frac{\sqrt{Kn}}{\hat{\sigma}_{(u,v)}^{(2)}}\left\langle u,\hat{L}^{(2)}v\right\rangle\xrightarrow{\text {d}} N(0,1),\quad \left\|\hat{R}^{(2)}\right\|_{F}=O_p\left(\frac{p}{Kn}\right). $$

\end{corollary}

Following the proof of Theorem \ref{theorem:2rdpca} and Corollary \ref{coro:eigenvalues}, we know that $\hat{\sigma}_{(u,v)}^{(t)}$ is a consistent distributed estimator of $\sigma_{(u,v)}$ for $t=2$; only in Algorithm \ref{alg:frdpca}, the local machines need to send $\tilde{\Sigma}_k \hat{U}^{(t)}$ and $(\tilde{\sigma}^{(t)}_{k})^{2}$ separately to the central machine in order to calculate $(\hat{\sigma}^{(t)})^2=\sum_{i=1}^{K}(\tilde{\sigma}^{(t)}_{k})^{2}/K$, instead of sending $\tilde{G}^{(t)}_k:=\tilde{\Sigma}_k \hat{U}^{(t)}-(\tilde{\sigma}^{(t)}_{k})^{2} \hat{U}^{(t)}$ as a whole, resulting in an additional $O(K)$ of communication. It is also possible to extend Corollary \ref{coro:an} to non-Gaussian cases using Proposition \ref{lem:lt}, but it involves a second term of asymptotic variance that is related to the fourth moment of the distribution and also knowledge on the unknown $\Sigma$, as depicted by the second term of \eqref{eq:a_variance}, which can be difficult to estimate from data \citep{koltchinskii2017new}. For statistical inference of the empirical spectral projectors in more general cases, \cite{naumov2019bootstrap,silin2020hypothesis} suggest using bootstrapping methods to perform hypothesis testing. We leave this interesting but challenging problem for future research.

\subsection{More General Cases}\label{sec:bias}

In this section, we work on more general cases, where $\{x_{i}^{k}\}$ are assumed to be sub-Gaussian; one can refer to \cite{vershynin2018high} for details of the sub-Gaussian distributions. In these cases, the one-round estimator is generally biased, unless the distribution is, e.g., symmetrically innovated \citep{fan2019distributed}.

\begin{assumption}[Sub-Gaussian distribution \citep{fan2019distributed}]\label{assump:2}
Assume that $x_{i}^{k}=A z_i^{k}\in\reals^p$, where $z_i^{k}$ consists of i.i.d. sub-Gaussian random variables with zero mean. The eigenvalues of the population covariance matrix $\Sigma=AA^T$ are $0\leq \lambda_p \leq \cdots\leq\lambda_{r+1}<\lambda_{r}\leq \cdots\leq\lambda_1<\infty$. Define $\sigma^2 := \sum_{j=r+1}^{p}\lambda_j/(p-r)$ and $\delta:=\max(|\lambda_{r+1}-\sigma^2|,|\lambda_{p}-\sigma^2|)$. Assume that for sufficiently large $K$, $n$ and $p$, we have $n\geq C_1p$ and $Kn\leq C_2 p^3$ for some constants $C_1, C_2>0$. Assume that $\lambda_{r}\geq\max(\lambda_{r+1}+d,\sigma^2+\delta+c_1)$ for some constants $d, c_1>0$.
\end{assumption}

The main part of Assumption \ref{assump:2} is taken directly from \citep{fan2019distributed}. For instance, $n\geq C_1p$ ensures a sufficiently large local signal-to-noise ratio, so that the one-round distributed estimator is applicable. This condition is required by Proposition \ref{prop:bias-fan}. It is in the same spirit as $l_r> \sqrt{c}$ required by Theorem \ref{theorem:1rdpca}, except that the latter is more accurate. Meanwhile, $\lambda_{r}\geq\sigma^2+\delta+c_1$ arises from the eigenvalue shifting from Algorithm \ref{alg:frdpca}, and the signal eigenvalues are required to lead after shifting. This condition is not necessary if we perform subspace iteration without eigenvalue shifts. However, as shown later by Theorem \ref{theorem:3} and Remark \ref{remark5}, the eigenvalue shifted version converges faster in many cases with weaker local SNR. Finally, $Kn\leq C_2 p^3$ is only to ensure a clearer expression of our results, under which the variance term $\sqrt{pr/Kn}$ is the leading term in our derivations. For a similar purpose, \cite{fan2019distributed} requires $Kp\leq C_3 n$ for some $C_3>0$ in their Theorem 4, which is more restrictive than ours when $p\asymp n$. For convenience, we restate Theorem 4 from \citep{fan2019distributed} as follows, using the notations of this work.

\begin{proposition}[Theorem 4 from \citep{fan2019distributed}]\label{prop:bias-fan} Under Assumption \ref{assump:2}, we have
\begin{equation}\label{eq:bound-fan}
    \left\|\hat{U}^{(1)}\left(\hat{U}^{(1)}\right)^{\top}-UU^{\top}\right\|_F=O_p\left(\underbrace{\sqrt{\frac{pr}{Kn}}}_{\text{Variance}}+\underbrace{\frac{\sqrt{r}p}{n}}_{\text{Bias}}\right).
\end{equation}
\end{proposition}

The right-hand side of \eqref{eq:bound-fan} consists of two terms. The variance $\sqrt{pr/Kn}$ and the bias $\sqrt{r}p/n$. Under weaker local SNR, i.e., $n\asymp p$ and $K\rightarrow \infty$, the bias term dominates the variance term. Even worse, the potential bias could be of constant order and $U^{(1)}$ is no longer consistent, leading to $\lim\limits_{k\rightarrow \infty} \lim\limits_{n,p\rightarrow \infty} \mathcal{V}(\hat{U}^{(1)}) =\infty$.  Fortunately, we show that the proposed method is able to alleviate the potential bias after an acceptable number of iterations. Recall that $\lceil a \rceil$ means the smallest integer that is not smaller than $a$, for any $a\in \reals$.

\begin{theorem}[Bias alleviation]\label{theorem:3}
Under Assumption \ref{assump:2}, for any $\hat{U}^{(1)}$ such that $\|\hat{U}^{(1)}(\hat{U}^{(1)})^{\top}-UU^{\top}\|_2<1$, in one of the following two cases:

1. If $\delta=O[(pr/Kn)^{\alpha}]$ for any $\alpha>0$,  set $T=\max(3,\lceil (2\alpha+1)/2\alpha +\varepsilon\rceil)$ for $\varepsilon>0$.

2. If $\delta \geq c_2$ for some $c_2>0$, set $T= \log^{1+\varepsilon} (Kn/pr)$ for some $\varepsilon>0$. 
$$\lim\limits_{k\rightarrow \infty} \lim\limits_{n,p\rightarrow \infty} \mathcal{V}\left(\hat{U}^{(T)}\right) =1.$$
\end{theorem}

In Theorem \ref{theorem:3}, the quantity $\delta$ controls the heterogeneity of the noise eigenvalues. If $\alpha$ in the case 1 is sufficiently large, then the noise eigenvalues are sufficiently uniform, and the eigenvalue shifting guarantees finite step convergence. For instance, if $\alpha> 1/4$, then $T=3$. Note that under Assumption \ref{assump:1}, the one-round estimator $\hat{U}^{(1)}$ is unbiased; hence the second round of communication suffices for statistical efficiency, as shown by Theorem \ref{theorem:2rdpca}. For more general cases where the bias could be of constant order, the third round is necessary. As $\alpha$ gets smaller, i.e., the noise eigenvalues are less uniform, more iteration rounds are needed.

If the noise eigenvalues are indeed heterogeneous, as depicted by the case $2$, then approximately $\log (K)$ communication rounds are required to alleviate the constant order bias, resulting in the total communication cost of $O(K\log K pr)$. It can still be much smaller than $O[Kp\min(p,n)]$, the communication cost of the pooling PCA, as long as $\log(K)\ll \min(p,n)$. Note that in Theorem \ref{theorem:3}, the requirement for the initialization is rather mild. Hence, the few-round estimator is expected to converge after a reasonable number of iterations even if the first-round estimator is questionable, say, under the local phase transition threshold; see  Figure \ref{fig-phase} and Figure \ref{fig:cases} for numerical validations.

\begin{remark}[Subspace iteration without eigenvalue shift]\label{remark5}
    It is totally feasible to perform subspace iteration without shifting the noise eigenvalues: after acquiring $\hat{U}^{(1)}$, the local machines only send $\tilde{\Sigma}_k\hat{U}^{(1)}$ to the central machine; and the central machine orthogonalizes $\sum_{k=1}^{K}\tilde{\Sigma}_k\hat{U}^{(1)}/K$ to acquire $\tilde{U}^{(2)}$; the procedure can be iterated until convergence.

    The preceding $\tilde{U}^{(t)}$ also converges to $\hat{U}^{\cP}$ in a linear convergence rate, but with the eigenvalue ratio $|\lambda_{r+1}/\lambda_{r}|$, instead of the eigenvalue ratio $|\delta/(\lambda_{r}-\sigma^2)|$ corresponding to the eigenvalue shifting version. In the local SNR regime of this work, $|\lambda_{r+1}/\lambda_{r}|$ shall be viewed as of constant order, and $\log(K)$ communication rounds are needed. Otherwise, the one-round estimator is already satisfactory and further subspace iteration is not necessary. Via eigenvalue shifting, we hopefully reduce the eigenvalue ratio and accelerate the convergence. In Table \ref{tab:results}, we compare the subspace iteration methods with ($\hat{U}^{(t)}$) or without ($\tilde{U}^{(t)}$) eigenvalue shift, and the eigenvalue-shifted version exhibits faster convergence.
\end{remark}

\subsection{Extension to distributed elliptical PCA}\label{sec:epca}
The previous sections discuss the statistical advantage of the few-round distributed PCA algorithm under sub-Gaussian distributions. This assumption might be restrictive in applications like finance and macroeconomics, where outliers or heavy-tailed data are common . Fortunately, our distributed PCA framework adapts readily to more general cases. In this section, we discuss the possible extension to distributed elliptical PCA using the multivariate Kendall's $\tau$ matrix, where moment constraints are not required \citep{Fan2018LARGE,han2018eca,he2022distributed}.

Elliptical family is a distribution family containing common distributions including Gaussian and $t$-distributions. It has nice properties, e.g.,  linear combinations of elliptically distributed random vectors also follow the elliptical distribution. We say a $p$-dimensional random vector $x_{i}^{k}$ is elliptically distributed with zero mean if and only if $x^{k}_i = r_{i}^{k} A u_{i}^{k}$, where $u_{i}^{k}$ is uniformly distributed on the unit sphere $S^{p-1}$ and is independent of the positive random radius $r_{i}^{k}$. Here $A\in\mathbb{R}^{p\times q_k}$ is a deterministic matrix satisfying $\Sigma=AA^{\top}$ and $\Sigma$ is called the scatter matrix of $x_{i}^{k}$. The statistical object of interest is the leading $r$ eigenspace of $\Sigma$, which is denoted by the column orthogonal matrix $U\in \reals^{p\times r}$ in this section.

Multivariate elliptical distribution is often used to model heavy-tailed data commonly observed in, e.g., financial markets \citep{han2018eca,Fan2018LARGE}. Elliptical PCA works on multivariate Kendall's $\tau$ matrix. The (local) sample version of the matrix is defined as follows:
$$\tilde{\Sigma}^{\tau}_{k}=\frac{2}{n(n-1)}\sum_{i<j}\frac{(x_{i}^{k}-x_{j}^{k})(x_{i}^{k}-x_{j}^{k})^{\top}}{\|x_{i}^{k}-x_{j}^{k}\|^2}.$$
We can plug $\tilde{\Sigma}^{\tau}_{k}$ into Algorithm \ref{alg:frdpca}, instead of the sample covariance matrix $\tilde{\Sigma}_k$ therein, so as to acquire robust distributed eigenspace estimation under potentially heavy-tailed distributions. We denote the resulting $t$-round estimator as $\hat{U}^{(t)}_{\tau}$; note that $\hat{U}^{(1)}_{\tau}$ is equivalent to the robust distributed estimator proposed by \cite{he2022distributed}.

\begin{assumption}[Elliptical spiked model \citep{he2022distributed}]\label{assum:ellip}
Assume that $x^{k}_i = r_{i}^{k} A u_{i}^{k}\in \reals^{p}$, where $u_{i}^{k}$ is uniformly distributed on the unit sphere $S^{p-1}$ and is independent of the positive random radius $r_{i}^{k}$. The eigenvalues of the population scatter matrix $\Sigma=AA^T$ are $0\leq \lambda_p = \cdots=\lambda_{r+1}<\lambda_{r}\leq \cdots\leq\lambda_1<\infty$. Assume that for sufficiently large $K$, $n$ and $p$, we have $Kn\leq C_4p^2\log p$ and $n \geq C_5p\log p$ for some constants $C_4, C_5>0$. Assume that $\lambda_r\geq \lambda_{r+1}+d$ and $\lambda_1/\lambda_p\leq C_6p^{\alpha}$ for some constants $d, C_6>0$ and fixed $0<\alpha<1/2$.
\end{assumption}

Most conditions are taken directly from \cite{he2022distributed}, and are similar to those in Assumption \ref{assump:2}. For example, $n \geq C_5p\log p$ is to ensure a sufficiently large local SNR, while $Kn\leq C_4p^2\log p$ leads to a leading variance term. Also note that the eigenvalues $\{\lambda_i\}$ of the scatter matrix are not identical to those $\{\lambda_i^{\tau}\}$ of the population Kendall's $\tau$ matrix given later in \eqref{eq:eg-kendall}. The homogeneity of $\{\lambda_i^{\tau}\}$ essentially relates to the convergence rate of the proposed algorithm. We assume homogeneous $\{\lambda_i\}$ and hence homogeneous $\{\lambda_i^{\tau}\}$ in Assumption \ref{assum:ellip}, mostly for presentational brevity. In the following, we restate Corollary 4.1 from \cite{he2022distributed}. 

\begin{proposition}[Corollary 4.1 from \cite{he2022distributed}]
    Under Assumption \ref{assum:ellip} we have
    \begin{equation}\label{eq:bound-he}
    \left\|\hat{U}^{(1)}_{\tau}\left(\hat{U}_{\tau}^{(1)}\right)^{\top}-UU^{\top}\right\|_F=O_p\left(\underbrace{\sqrt{\frac{p\log p r^3}{Kn}}}_{\text{Variance}}+\underbrace{\frac{r^{5/2}p^2\log p}{n}}_{\text{Bias}}\right).
\end{equation}
\end{proposition}

Similar to Proposition \ref{prop:bias-fan}, the potential bias term can be the leading term under weaker local SNR. Fortunately, as shown by the following Corollary \ref{theorem:ellip}, the proposed method alleviates the bias after further consensus rounds. Meanwhile, the variance term is also slightly smaller than the one in \eqref{eq:bound-he} by a factor of $r$.

\begin{corollary}[Few-round distributed elliptical PCA]\label{theorem:ellip}
    Under Assumption \ref{assum:ellip}, for any $\hat{U}_{\tau
    }^{(1)}$ such that $\|\hat{U}_{\tau
    }^{(1)}(\hat{U}_{\tau
    }^{(1)})^{\top}-UU^{\top}\|_2<1$, we have
    \begin{equation}\label{eq:ba-ellip}
    \left\|\hat{U}^{(3)}_{\tau}\left(\hat{U}_{\tau}^{(3)}\right)^{\top}-UU^{\top}\right\|_F=O_p\left(\sqrt{\frac{p\log p r}{Kn}}\right).
\end{equation}
\end{corollary}

\section{Numerical Experiments}
In this section, we conduct numerical experiments on synthetic datasets and benchmark tabular datasets to validate our theoretical claims and also demonstrate the empirical advantage of the proposed few-round method.\footnote{One can find the codes to reproduce the numerical results at \url{https://github.com/LZeY-FD/FRDPCA}.}     

\subsection{Synthetic Datasets}\label{sec:synds}

\begin{figure}[h]
	\centering
    \begin{minipage}{0.315\linewidth}
		\centering
	\includegraphics[width=0.88\linewidth]{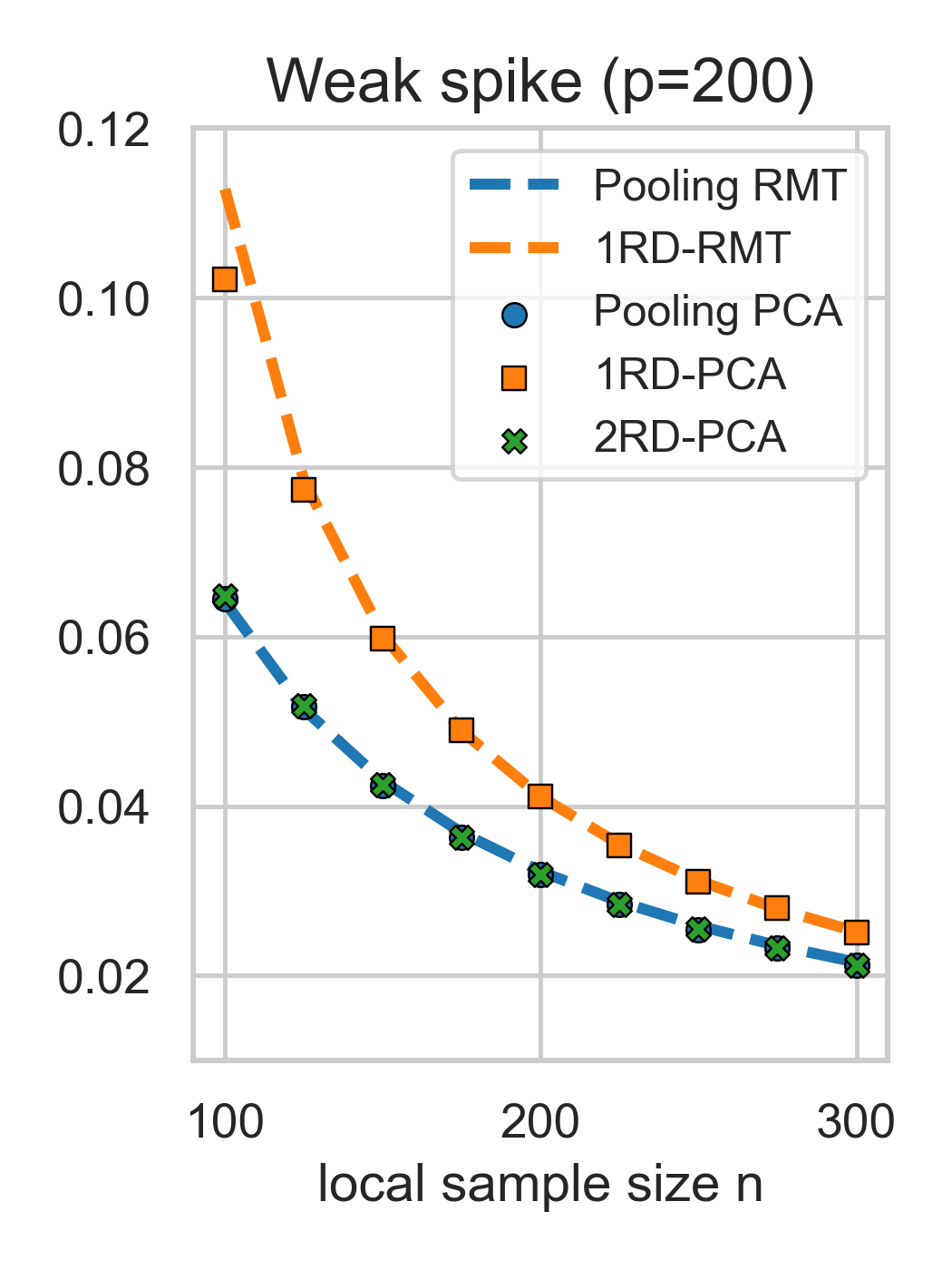}
	\end{minipage}
        \begin{minipage}{0.315\linewidth}
		\centering
	\includegraphics[width=0.88\linewidth]{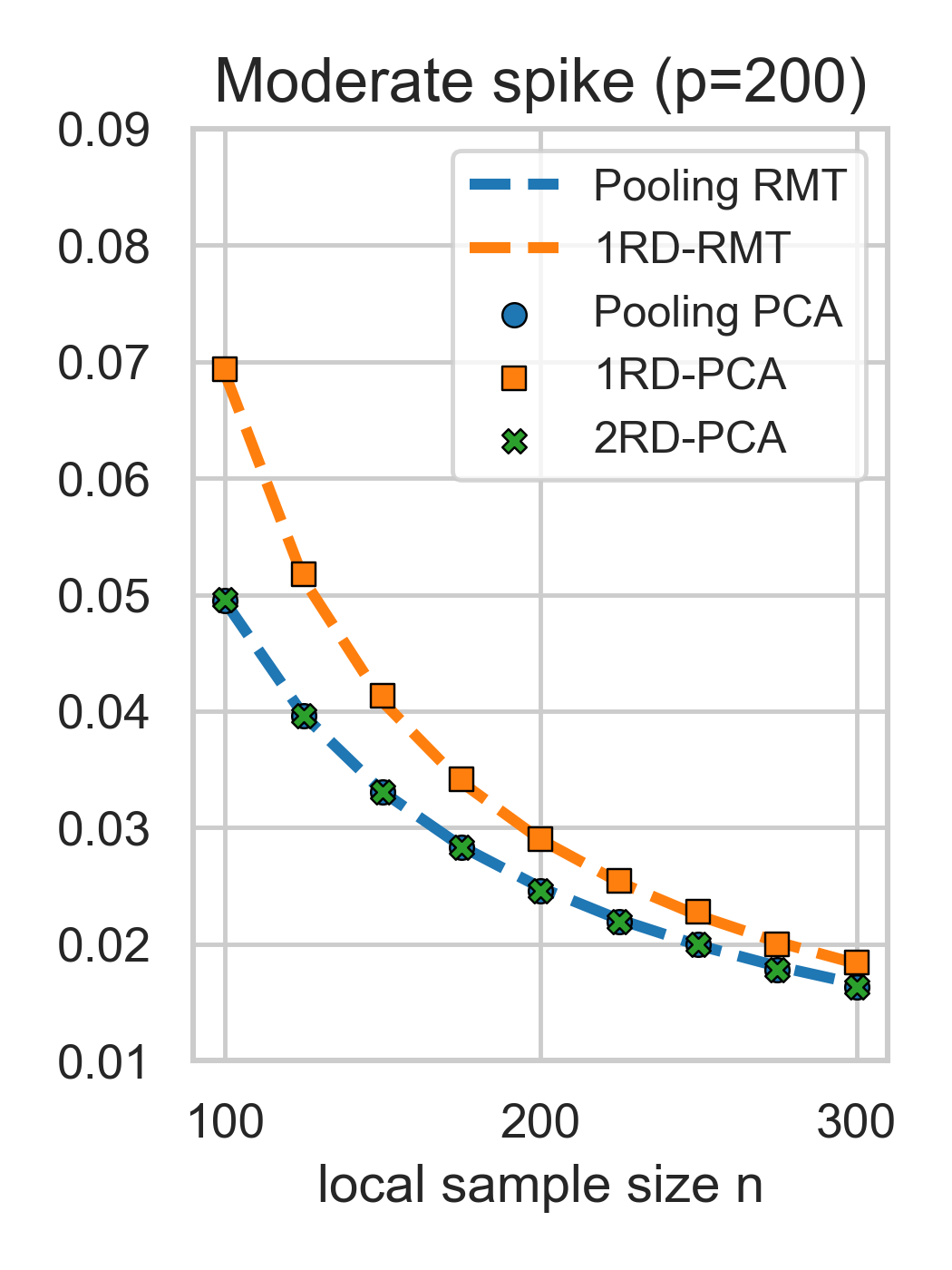}
	\end{minipage}
	\begin{minipage}{0.315\linewidth}
		\centering
	\includegraphics[width=0.88\linewidth]{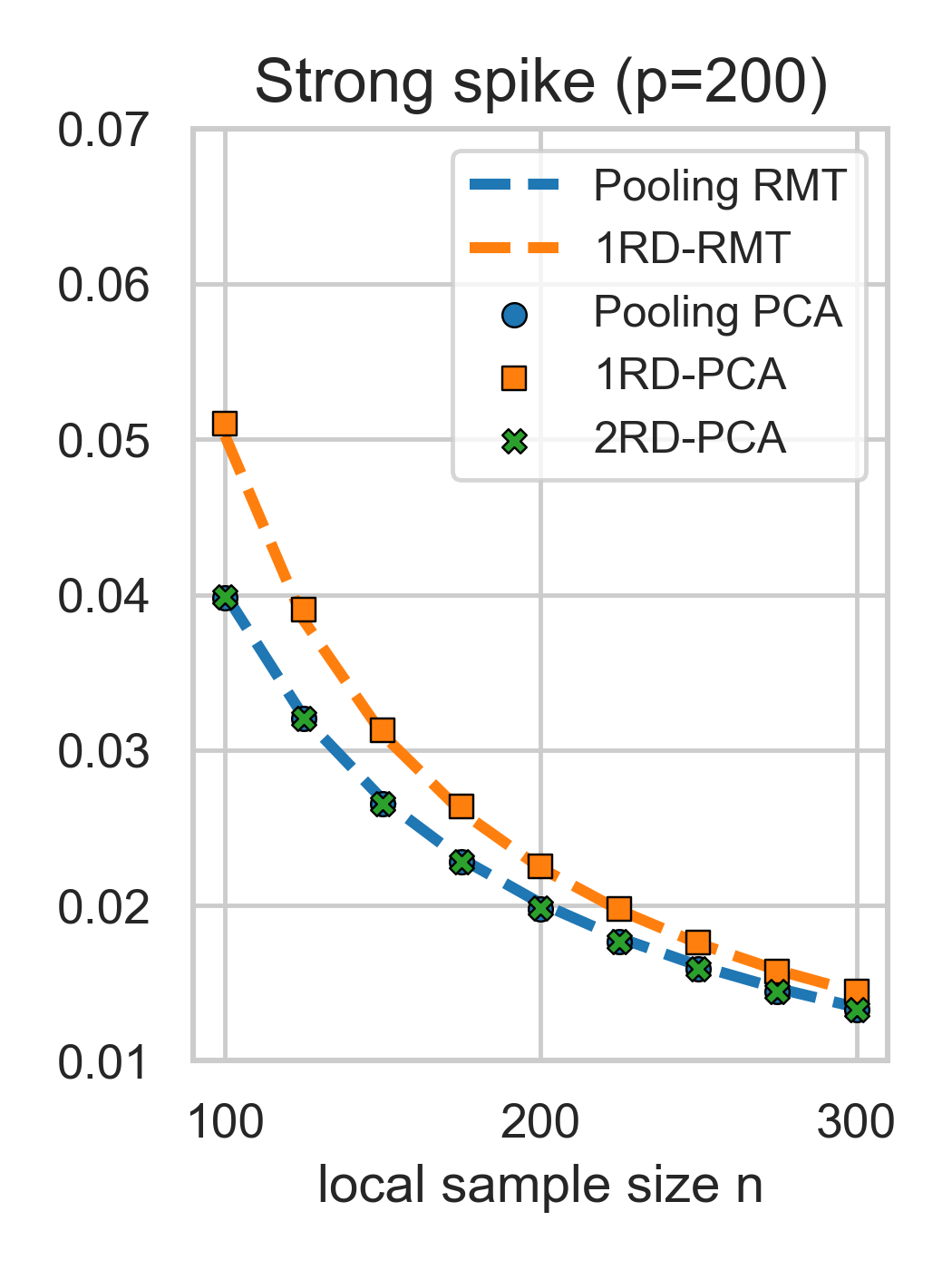}
	\end{minipage}
\caption{Mean squared subspace estimation error of different estimators on synthetic datasets (depicted by orange, blue and green dots), where the one-round estimator is unbiased, with respect to relatively weak spike, moderate spike and relatively strong spike. The results are based on $100$ replications for $p=200$, and $n$ ranges from $100$ to $300$. The numerical results are plotted against the theoretical rates (depicted by orange and blue dashed lines) derived from the random matrix theory, i.e., Lemma \ref{lemma:pool} and Lemma \ref{lemma:1rdpca}.\label{fig-sim-main}}
\end{figure}

We generate synthetic datasets to validate our theoretical arguments. First, following Assumption \ref{assump:1}, we generate independent centered Gaussian random vectors with spike strength $$(l_1,l_2,l_3)\in\{(2.5,2.25,2),(3,2.75,2.5),(3.5,3.25,3)\},$$
corresponding to three scenarios: relatively weak spike, moderate spike, and relatively strong spike. In these cases, the one-round estimator is unbiased. We set $p =200$ and $K = 60$, while $n$ ranges from $100$ to $300$. In Figure \ref{fig-sim-main}, we report the mean squared subspace estimation error, as measured by $\|\cdot-UU^{\top}\|_F^2/2$, of different estimators based on $100$ replications.

We can see that the mean squared subspace estimation error of the pooling PCA estimator and the one-round distributed PCA estimator matches the theoretical rates derived from random matrix theory, i.e., Lemma \ref{lemma:pool} and Lemma \ref{lemma:1rdpca}. As predicted by Theorem \ref{theorem:1rdpca}, the one-round distributed PCA estimator tends to be more efficient as the local signal-to-noise ratio increases, i.e., either $n$ or $l_i$ gets larger. However, the one-round estimator still has a larger statistical error than the pooling PCA estimator under relatively weaker local SNR. Meanwhile, the two-round distributed PCA estimator manages to achieve similar statistical performance as the pooling PCA estimator.

For the more general cases, our data generation process considers both symmetric and asymmetric distributions with varying noise structures. For symmetric distributions, we generate centered multivariate Gaussian distributions. As for asymmetric distributions, we consider skew-Gaussian distributions with skewness parameter $\alpha=5$ for signal dimensions, and $\alpha=2$ for noise dimensions, using the \texttt{Python} package \texttt{scipy.stats}. The one-round estimator is generally biased under asymmetric distributions. Meanwhile, under uniform noise, all noise eigenvalues are set to $1$. For decaying noise, the noise eigenvalues linearly decrease from $1.2$ to $0.8$. The population covariance matrix has eigenvalues $(6,4,3)$ for the signal components. We set the dimension $p=200$ and distribute the data across $K=60$ machines, with $n=200$ samples per machine, and $T$ ranges from $1$ to $3$. We compare subspace iteration estimators with and without eigenvalue shift, denoted as $\hat{U}^{(t)}$ and $\tilde{U}^{(t)}$, respectively. All results are averaged over $100$ independent replications.

\begin{table}[htbp]
\centering
\caption{Comparison of few-round distributed PCA estimators across different scenarios, in terms of the Frobenius norm, based on $100$ replications. Note that the one-round estimator is unbiased under symmetric distributions, but is generally biased under asymmetric distributions. The subspace iterated estimators with eigenvalue shift are denoted by $\hat{U}^{(t)}$, while the ones without eigenvalue shift are denoted by $\tilde{U}^{(t)}$. Note that $\hat{U}^{(1)}$ is equivalent to the one-round estimator from \cite{fan2019distributed}.}
\label{tab:results}
\resizebox{0.9\textwidth}{!}{%
\begin{tabular}{ccccc}
\toprule
Gaussian  & Sym. Uniform & Asym. Uniform & Sym. Decaying & Asym. Decaying \\
\midrule
$\hat{U}^{(1)}$ & 0.0293 $\pm$ 0.0020 & 0.0629 $\pm$ 0.0048 & 0.0302 $\pm$ 0.0021 & 0.0663 $\pm$ 0.0064 \\
$\tilde{U}^{(2)}$ & 0.0238 $\pm$ 0.0016 & 0.0408 $\pm$ 0.0032 & 0.0245 $\pm$ 0.0016 & 0.0442 $\pm$ 0.0037 \\
$\tilde{U}^{(3)}$ & 0.0234 $\pm$ 0.0013 & 0.0383 $\pm$ 0.0028 & 0.0240 $\pm$ 0.0016 & 0.0397 $\pm$ 0.0033 \\
$\hat{U}^{(2)}$ & 0.0234 $\pm$ 0.0013 & 0.0379 $\pm$ 0.0026 & 0.0239 $\pm$ 0.0016 & 0.0389 $\pm$ 0.0030 \\
$\hat{U}^{(3)}$ & 0.0234 $\pm$ 0.0013 & 0.0377 $\pm$ 0.0026 & 0.0239 $\pm$ 0.0016 & 0.0384 $\pm$ 0.0030 \\
$\hat{U}^{\cP}$ & 0.0234 $\pm$ 0.0013 & 0.0377 $\pm$ 0.0026 & 0.0239 $\pm$ 0.0016 & 0.0384 $\pm$ 0.0030 \\
\bottomrule
\end{tabular}
}
\end{table}

In Table \ref{tab:results}, we report the comparison of estimators across different scenarios, in terms of the subspace Frobenius norm, e.g., $\|\hat{U}^{(t)}(\hat{U}^{(t)})^{\top}-UU^{\top}\|_F$.  We can see that the proposed few-round distributed PCA works well under all cases. It is also observed that $\hat{U}^{(2)}$ is almost as good as $\hat{U}^{\cP}$, except for the cases with asymmetric distributions, where the one-round estimator is biased and the third communication round is suggested.

\subsection{Benchmark tabular datasets}
We apply the distributed PCA algorithms to the benchmark tabular datasets from \cite{grinsztajn2022tree}. In \cite{grinsztajn2022tree}, the authors compiled 45 tabular datasets from various domains, provided mostly by OpenML. The benchmark consists of four splits based on tasks and datasets used for tasks, which are respectively regression on numerical features, regression on numerical and categorical features, classification on numerical features, and classification on categorical features.\footnote{The benchmark datasets and the corresponding details could be found at \url{https://huggingface.co/datasets/inria-soda/tabular-benchmark}.} Thus, we are able to experiment with the distributed PCA algorithms on numerical and categorical features of these datasets. We discard the datasets with small feature dimensions ($p<30$), resulting in $10$ datasets left; see Table \ref{tab:Np} for basic dimensional information on the remaining datasets.

Each remaining benchmark dataset is viewed as a pooled dataset. We randomly take $80\%$ of each benchmark dataset as the training set and set the remaining $20\%$ as the test set. Given constants $\kappa$ and $\rho$, we split the pooled training dataset with local sample sizes $n=\lfloor\kappa p \rfloor$. In Table \ref{tab:results_table_benchmark}, we take at most $K=1000$ local machines to save computation. We apply distributed PCA algorithms with cut-off dimension $r=\min(\lfloor \rho p \rfloor, r_{\max})$ for some pre-determined $r_{\max}$. In our experiments, we set $\kappa=1$, $\rho=0.1$, and $r_{\max}=5$.

In Table \ref{tab:results_table_benchmark}, we report the relative performance of different distributed PCA estimators against the pooling PCA estimator. The statistical performance for any estimator $\hat{U}$ is quantified by the test set Average information preservation Ratio (AR), defined as 
$$\text{AR}(\hat{U})=\frac{\sum_{i}\|\hat{U}\hat{U}^{\top}x^{\text{test}}_i\|_2^2}{\sum_{i}\|x^{\text{test}}_i\|_2^2},$$ 
where $\|\cdot\|_2$ is the vector $\ell_2$ norm. We consider subspace iteration estimators with ($\hat{U}^{(t)}$) or without ($\tilde{U}^{(t)}$) eigenvalue shift, and also the SIP estimator ($\hat{U}^{(T\times T')}_{\text{SIP}}$) with $T\in\{5,10\}$, $T'=5$ and $\eta=1$ from \cite{chen2022distributed}. Note that $\hat{U}^{(1)}$ is equivalent to the one-round estimator from \cite{fan2019distributed}.

\begin{table}
\centering
\caption{The average relative performance (and computation time) of different distributed PCA estimators against the pooling PCA estimator, over 10 selected benchmark tabular datasets, based on $100$ replications. We compare subspace iteration estimators with ($\hat{U}^{(t)}$) or without ($\tilde{U}^{(t)}$) eigenvalue shift, and the SIP estimator ($\hat{U}^{(T\times T')}_{\text{SIP}}$) with $T\in\{5,10\}$, $T'=5$ and $\eta=1$ from \cite{chen2022distributed}. Note that $\hat{U}^{(1)}$ is equivalent to the one-round estimator from \cite{fan2019distributed}.}
\label{tab:results_table_benchmark}
\resizebox{0.9\textwidth}{!}{%
\begin{tabular}{lccccc}
\toprule
Method & road-safety & MiniBooNE & jannis & Ailerons & yprop\_4\_1 \\
\midrule
$\hat{U}^{(1)}$ & 0.931 (0.166s) & 0.145 (0.217s) & 0.992 (0.210s) & 0.999 (0.052s) & 1.004 (0.033s) \\
$\tilde{U}^{(2)}$ & 0.971 (0.262s) & 1.007 (0.325s) & 0.995 (0.306s) & 1.000 (0.082s) & 1.011 (0.049s) \\
$\tilde{U}^{(3)}$ & 0.989 (0.270s) & 1.011 (0.333s) & 0.996 (0.309s) & 1.000 (0.083s) & 1.010 (0.050s) \\
$\hat{U}^{(2)}$ & 0.980 (0.367s) & 0.998 (0.438s) & 0.996 (0.403s) & 1.000 (0.112s) & 1.011 (0.065s) \\
$\hat{U}^{(3)}$ & 0.995 (0.467s) & 1.005 (0.558s) & 0.998 (0.501s) & 1.000 (0.146s) & 1.010 (0.085s) \\
$\hat{U}^{(5\times5)}_{\text{SIP}}$ & 0.783 (3.946s) & 1.008 (6.806s) & 0.839 (5.852s) & 0.951 (1.269s) & 0.885 (0.897s) \\
$\hat{U}^{(10\times5)}_{\text{SIP}}$ & 0.794 (6.921s) & 1.000 (11.894s) & 0.845 (10.215s) & 0.955 (2.169s) & 0.894 (1.537s) \\
\midrule
Method & Bioresponse & Allstate & Mercedes\_Benz & topo\_2\_1 & superconduct \\
\midrule
$\hat{U}^{(1)}$ & 0.989 (0.061s) & 0.991 (0.862s) & 0.983 (0.055s) & 1.000 (0.090s) & 1.000 (0.080s) \\
$\tilde{U}^{(2)}$ & 0.999 (0.077s) & 0.999 (1.073s) & 0.999 (0.067s) & 1.001 (0.106s) & 1.000 (0.109s) \\
$\tilde{U}^{(3)}$ & 0.999 (0.084s) & 0.999 (1.083s) & 1.001 (0.072s) & 1.000 (0.109s) & 1.000 (0.110s) \\
$\hat{U}^{(2)}$ & 0.999 (0.092s) & 0.999 (1.318s) & 0.999 (0.079s) & 1.000 (0.125s) & 1.000 (0.138s) \\
$\hat{U}^{(3)}$ & 0.999 (0.114s) & 1.000 (1.578s) & 1.001 (0.095s) & 1.000 (0.145s) & 1.000 (0.170s) \\
$\hat{U}^{(5\times5)}_{\text{SIP}}$ & 0.902 (0.327s) & 0.648 (8.818s) & 0.685 (0.265s) & 0.945 (0.393s) & 0.962 (1.503s) \\
$\hat{U}^{(10\times5)}_{\text{SIP}}$ & 0.903 (0.494s) & 0.647 (14.955s) & 0.684 (0.418s) & 0.945 (0.626s) & 0.964 (2.529s) \\
\bottomrule
\end{tabular}
}
\end{table}

While in most cases, the one-round distributed estimator achieves satisfactory accuracy (due to relatively high local SNR), we still see that the proposed few-round distributed PCA algorithm shows a persistent statistical advantage against the one-round distributed PCA algorithm in all $10$ datasets. In particular, for the MiniBooNE dataset, the one-round distributed PCA breaks down and fails to retrieve the population principal subspace. Meanwhile, the few-round distributed PCA is still robust even though the one-round subspace estimator is no longer valid. It is consistent with the phase transition gap discussed earlier in Table \ref{tab:method_comparison} and also shown in Figure \ref{fig-phase}: if we over-split the dataset, the local signal-to-noise ratio can be so small that the local subspace estimators are questionable. In conclusion, the further consensus rounds can increase the statistical accuracy. We also conducted experiments with additional communication rounds and other sets of $(\kappa, \rho,r_{\max})$, but the results are almost identical and thus omitted. As a result, we gladly suggest adding a few, typically one to four, shifted subspace iteration rounds after acquiring the one-round estimator.

\begin{figure}[h]
  \centering
    \begin{minipage}[t]{1\linewidth}
  \centering
  \includegraphics[width=5in]{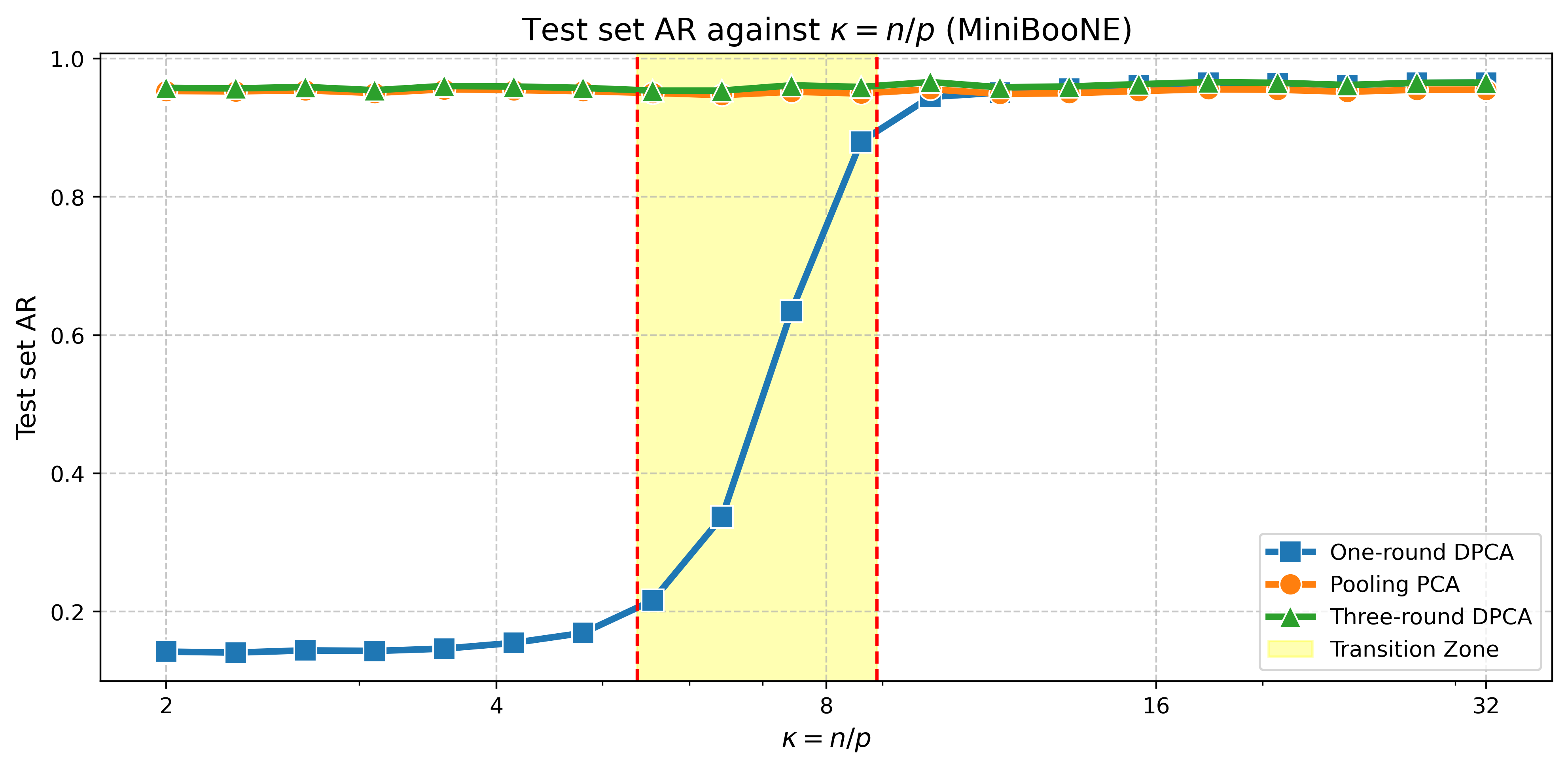}
  \includegraphics[width=5in]{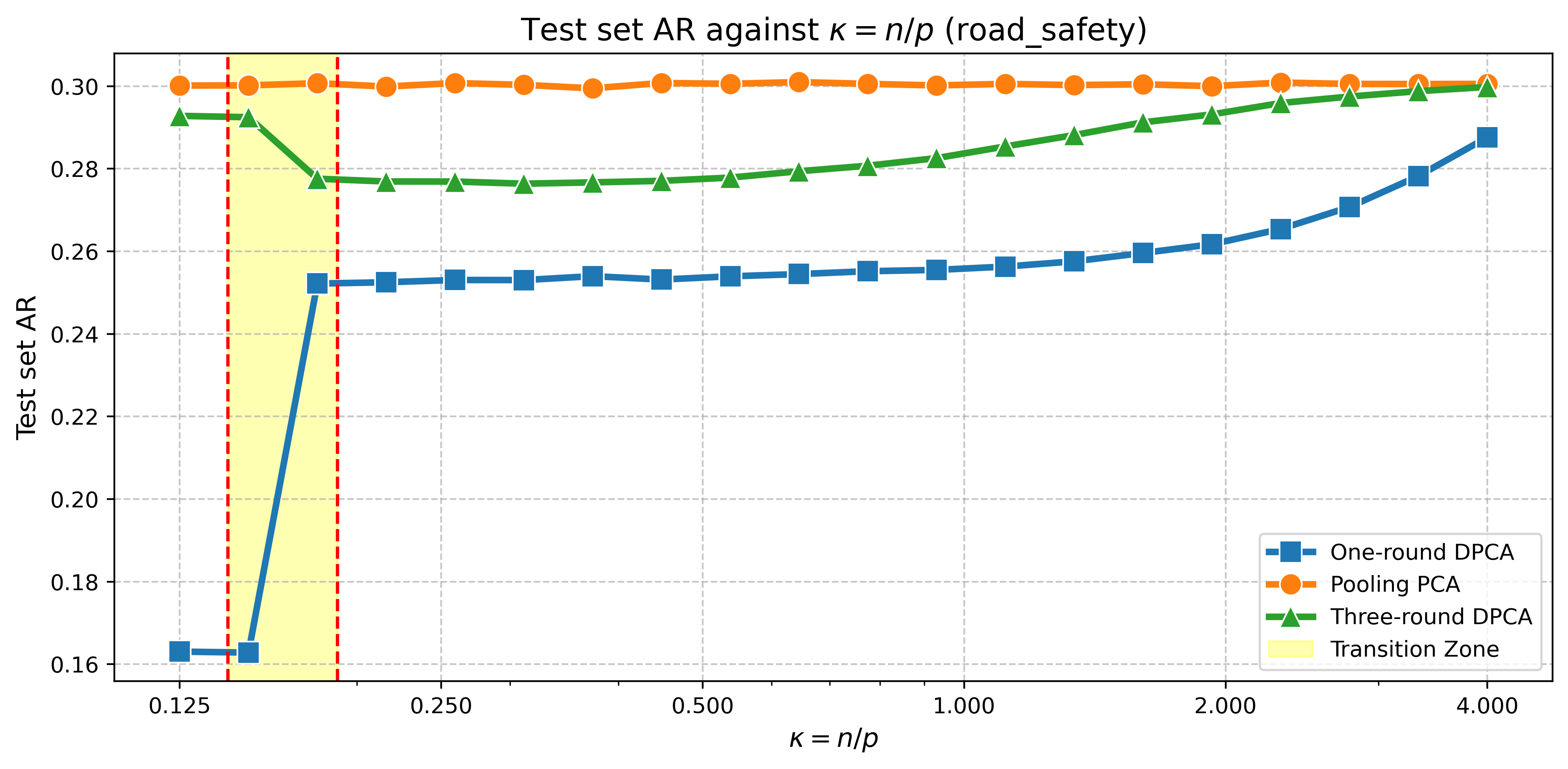}
  \end{minipage}
  \caption{In both MiniBooNE and road\_safety datasets, as the local SNR increases, the divide-and-conquer estimator becomes more statistically efficient. Also, we observe the local phase transition gaps due to over-splitting the pooled dataset. As the local machines have smaller local sample sizes and approaches the transition zone, the local subspace estimators breaks down rapidly and the one-round estimator turns questionable. Meanwhile, the few-round distributed estimator by consensus can close this gap by sharing the jointly estimated subspace from the previous round. The results are based on $100$ replications.}\label{fig:cases}
\end{figure}

Finally, we dive deeper into the two datasets where the one-round algorithm loses statistical efficiency due to over-splitting of the dataset, i.e., MiniBooNE and road\_safety. We randomly split the pooled datasets into local datasets with $n=\lfloor\kappa p \rfloor$, and vary $\kappa$ at logarithmic scales. Larger $\kappa$ means larger local sample size and hence higher local SNR. In Figure \ref{fig:cases}, we report the test set AR of different methods, based on $100$ replications. We can see that the divide-and-conquer estimator becomes more statistically efficient as the local SNR increases, which is consistent with the previous theoretical claims.

In both the MiniBooNE and road\_safety datasets, phase transition gaps caused by over-splitting of the pooled dataset are observed. As the local machines have smaller sample sizes and approach the transition regime, the local subspace estimators degrade rapidly, rendering the divide-and-conquer estimator unreliable. In contrast, the few-round distributed estimator based on consensus is able to bridge this gap by iteratively sharing and refining the estimated subspace across rounds.

\section{Discussion}\label{sec:dis}

In this paper, we show that the celebrated divide-and-conquer method for distributed PCA loses statistical efficiency under weaker local SNR, due to the local phase transition gap, the inflated variance, and the inherent bias. Fortunately, a few additional shifted subspace iteration steps can close the statistical efficiency gap by consensus. In our experiments on synthetic and benchmark tabular datasets, the proposed algorithm shows a persistent statistical advantage over the one-round algorithm.

For future work, it is possible to dive deeper into other cases where a few more steps of iteration are sufficient for statistical efficiency. Indeed, the iterative projection technique is popular in subspace estimation for higher-order data \citep{xia2021statistically}. It is intriguing to explore the common aspects of these problems in a unified manner. On the other hand, distributed estimation of matrix or tensor-valued parameters in, e.g., low-rank trace regression, is also a topic of both theoretical and practical interest.




\appendix

\section{Proof of Theoretical Results}\label{app:theorem}

Here we present the proof of our theoretical results.
\subsection{Proof of Theorem \ref{theorem:1rdpca}}

Before diving into the theoretical details, we first give a brief outline of the proof. The first step is to understand the asymptotic behavior of the pooling PCA estimator $\hat{U}^{\cP}$, which is rather straightforward thanks to the existing random matrix theory \citep{paul2007asymptotics,couillet2022random}. We know that the following Lemma \ref{lemma:pool} holds.

\begin{lemma}[Pooling PCA estimator]\label{lemma:pool}
    Under Assumption \ref{assump:1}, we have
    $$K \E\left(\left\|\hat{U}^{\cP}(\hat{U}^{\cP})^{\top}-UU^{\top}\right\|^2_F\right)\rightarrow 2\sum_{i=1}^{r}  \frac{pl_i^{-1}+p l_i^{-2}}{n}.$$
\end{lemma}

The second step is to study the asymptotic convergence of the one-round distributed estimator. First, as also shown in \cite{fan2019distributed}, the one-round distributed PCA is unbiased under Assumption \ref{assump:1}, in the sense that $\span(U)$ is also the leading eigenspace of $\Sigma^{(1)}:=\E(\tilde{U}_k\tilde{U}^{\top}_k)$. Then, using the matrix concentration inequality \citep{tropp2012user,tropp2015introduction}, we show that $\|\sum_{k=1}^{K}\tilde{U}_k\tilde{U}^{\top}_k/K-\E(\tilde{U}_k\tilde{U}^{\top}_k)\|_{2}$ tends to $0$ as $K$, $n$, $p\rightarrow\infty$. After that, we use the higher-order Davis-Kahan theorem from \cite{fan2019distributed} to carefully separate the linear matrix perturbation error from the higher-order term. The rate in the following Lemma \ref{lemma:1rdpca} could be obtained by calculating the squared Frobenius norm of this linear matrix perturbation term.

\begin{lemma}[One-round distributed PCA]\label{lemma:1rdpca}
    Under Assumption \ref{assump:1}, we have
    $$K\E \left(\left\|\hat{U}^{(1)}(\hat{U}^{(1)})^{\top}-UU^{\top}\right\|^2_F\right)\rightarrow 2\sum_{i=1}^{r}\frac{pl_i^{-1}+pl_i^{-2}}{n-pl_i^{-2}}.$$
\end{lemma}

Theorem \ref{theorem:1rdpca} is proved by combining both Lemma \ref{lemma:pool} and Lemma \ref{lemma:1rdpca}. We now present the proof of these lemmas.

\subsubsection{Proof of Lemma \ref{lemma:pool}}
By the relationship between $\|\hat{U}^{\cP}(\hat{U}^{\cP})^{\top}-UU^{\top}\|_F$ and $\tr(\hat{U}^{\cP}(\hat{U}^{\cP})^{\top}UU^{\top})$, as stated by Lemma \ref{lemma:subspace-error}, we focus on
\begin{equation*}
    \begin{aligned}
        \tr\left(\hat{U}^{\cP}(\hat{U}^{\cP})^{\top}UU^{\top}\right)&=\tr\left[\left(\sum_{i=1}^r \hat{u}_i^{\cP}(\hat{u}_i^{\cP})^{\top}\right)\left(\sum_{j=1}^r u_ju_j^{\top}\right)\right]\\
        &= \sum_{i=1}^r\sum_{j=1}^r \tr\left(\hat{u}_i^{\cP}(\hat{u}_i^{\cP})^{\top}u_ju_j^{\top}\right)= \sum_{i=1}^r\sum_{j=1}^r u_j^{\top}\hat{u}_i^{\cP}(\hat{u}_i^{\cP})^{\top}u_j.
    \end{aligned}
\end{equation*}

By Proposition \ref{prop:rmt-eigenvector}, for finite $K$ we have
$$u_i^{\top}\hat{u}_i^{\cP}(\hat{u}_i^{\cP})^{\top}u_i\xrightarrow{\text {a.s.}} \frac{Kn-p l_i^{-2}}{Kn+ pl_i^{-1}} 1_{l_i>\sqrt{p/Kn}},$$
$$u_j^{\top}\hat{u}_i^{\cP}(\hat{u}_i^{\cP})^{\top}u_j\xrightarrow{\text {a.s.}} 0,\quad i\neq j.$$

Hence, we have
$$\frac{1}{2}\left\|\hat{U}^{\cP}(\hat{U}^{\cP})^{\top}-UU^{\top}\right\|^2_F = r- \tr\left(\hat{U}^{\cP}(\hat{U}^{\cP})^{\top}UU^{\top}\right)= \sum_{i=1}^{r} \left(1 - u_i^{\top}\hat{u}_i^{\cP}(\hat{u}_i^{\cP})^{\top}u_i\right),$$

\begin{equation*}
    \begin{aligned}
        \frac{1}{2}\left\|\hat{U}^{\cP}(\hat{U}^{\cP})^{\top}-UU^{\top}\right\|^2_F&\xrightarrow{\text {a.s.}}\sum_{i=1}^{r} \left(1 - \frac{Kn-p l_i^{-2}}{Kn+ pl_i^{-1}} 1_{l_i>\sqrt{p/Kn}}\right)\\
        &= \sum_{i=1}^{r} \left( \frac{pl_i^{-1}+p l_i^{-2}}{Kn+ pl_i^{-1}}1_{l_i>\sqrt{p/Kn}}+1_{l_i\leq\sqrt{p/Kn}}\right).
    \end{aligned}
\end{equation*}
The lemma clearly holds as $K\rightarrow \infty$.

\subsubsection{Proof of Lemma \ref{lemma:1rdpca}}
First, we focus on the theoretical properties of $\Sigma^{(1)}=\E(\tilde{U}_k\tilde{U}^{\top}_k)$. In essence, we show that the leading $r$ eigenvectors of $\Sigma$ are also the leading $r$ eigenvectors of $\Sigma^{(1)}$.
\begin{lemma}[Expected projection matrix]\label{lem:epm}
    Under Assumption \ref{assump:1}, for $i\in [r]$, the $i$-th eigenvector of $\Sigma$, denoted as $u_i$, is also the $i$-th eigenvector of $\Sigma^{(1)}$. It corresponds to the eigenvalue
    $$\lambda^{(1)}_i\rightarrow \frac{n-pl_i^{-2}}{n+pl_i^{-1}}.$$
    Meanwhile, for $j\geq r+1$, we have $\lambda_j^{(1)}=\lambda_{r+1}^{(1)}\rightarrow 0$.
\end{lemma}

\begin{proof}
    On the $k$-th machine, we have $X_k=\Sigma^{1/2}Z_k\in \reals^{p\times n}$, where $Z_k\in \reals^{p\times n}$ consists of independent standard Gaussian random variables. For $u_i$, the $i$-th eigenvector of $\Sigma$, define the Householder matrix $H_i = (2u_iu_i^{\top}-I_p)$. Clearly, the reflection matrix $H_i$ and $\Sigma^{1/2}$ share the same set of eigenvectors, and hence are exchangeable. Thanks to the orthogonal invariance of Gaussian distribution, we have $$X'_k:=H_i\Sigma^{1/2}Z_k=\Sigma^{1/2}H_iZ_k\stackrel{d}{=}\Sigma^{1/2}Z_k=X_k.$$
    Let $\tilde{U}'_k:=H_i\tilde{U}_k$ be the leading $r$ eigenvectors of $\tilde{\Sigma}'_k = X_kX_k^{\top}/K$. Clearly, $\tilde{U}'_k$ has the same distribution as $\tilde{U}_k$. We have
    $$2\Sigma^{(1)}=2\E(\tilde{U}_k\tilde{U}^{\top}_k) = \E\left[\tilde{U}_k\tilde{U}^{\top}_k+ \tilde{U}'_k(\tilde{U}'_k)^{\top}\right].$$
Meanwhile, as $H_iu_i = u_i$, we have
 \begin{equation*}
     \begin{aligned}
     \left[\tilde{U}'_k(\tilde{U}'_k)^{\top}+\tilde{U}_k\tilde{U}_k^{\top}\right]u_i&=\left[(2u_iu_i^{\top}-I_p) \tilde{U}_k\tilde{U}_k^{\top} + \tilde{U}_k\tilde{U}_k^{\top}\right]u_i \\
         &= 2u_iu_i^{\top}\tilde{U}_k\tilde{U}_k^{\top} u_i = 2\left( u_i^{\top}\tilde{U}_k\tilde{U}_k^{\top} u_i\right)u_i.
     \end{aligned}
 \end{equation*}
 Hence, $u_i$ is still an eigenvector of $\Sigma^{(1)}$. Meanwhile, when $l_r>\sqrt{p/n}$ as $n$, $p\rightarrow \infty$, according to Proposition \ref{prop:rmt-eigenvector}, we have $\lambda_i^{(1)}\rightarrow (n-pl_i^{-2})/(n+pl_i^{-1})$.

 Then, we proceed to $j\geq r+1$. The symmetry $\lambda_j^{(1)}=\lambda_{r+1}^{(1)}$ comes from the fact that $\Sigma^{(1)}$ is exchangeable with
 $$H_j =\left(U,U_{\perp}\right) \left(\begin{array}{cc}I_{r} & 0_{r\times(p-r)} \\ 0_{(p-r)\times r} & A_j\end{array}\right)\left(\begin{array}{c}U^{\top} \\ U_{\perp}^{\top} \end{array}\right),$$
for any orthogonal matrix $A_j$. Hence, $\lambda_j^{(1)}=\lambda_{r+1}^{(1)}=(r-\sum_{i=1}^{r}\lambda_i^{(1)})/(p-r)\rightarrow 0$ as $n$, $p\rightarrow \infty$. The proof is complete.
\end{proof}

With Lemma \ref{lem:epm} in hand, we proceed to control $\|\sum_{k=1}^{K}\tilde{U}_k\tilde{U}^{\top}_k/K-\E(\tilde{U}_k\tilde{U}^{\top}_k)\|_{2}$ using matrix concentration inequality \citep{tropp2012user,tropp2015introduction}.

\begin{lemma}[Matrix concentration]\label{lem:mc}
Under Assumption \ref{assump:1}, we have
    $$\left\|\sum_{k=1}^{K}\tilde{U}_k\tilde{U}^{\top}_k/K-\E(\tilde{U}_k\tilde{U}^{\top}_k)\right\|_{2} = O_p(K^{-1/2})=o_p(1).$$
\end{lemma}

\begin{proof}
A direct application of matrix Hoeffding's inequality \citep{tropp2012user} yields the convergence rate of $O_p[(\log p/K)^{1/2}]$. Fortunately, in the case of random projection matrices, we are able to remove the dimensional factor, i.e., $\log p$, via intrinsic dimension arguments, see Theorem 7.7.1 in \citep{tropp2015introduction}.

We now verify the conditions of Theorem 7.7.1 in \citep{tropp2015introduction}. Clearly, $\tilde{U}_k\tilde{U}^{\top}_k-\E(\tilde{U}_k\tilde{U}^{\top}_k)$ is centered, and $\|\tilde{U}_k\tilde{U}^{\top}_k-\E(\tilde{U}_k\tilde{U}^{\top}_k)\|_{2}$ is bounded. For the matrix-valued variance $\tilde{V}$, we have
$$\tilde{V}/K=\E\left[\tilde{U}_k\tilde{U}^{\top}_k-\E(\tilde{U}_k\tilde{U}^{\top}_k)\right]^2=\E\left(\tilde{U}_k\tilde{U}^{\top}_k\right)-\E\left(\tilde{U}_k\tilde{U}^{\top}_k\right)^2,$$
whose eigenvalues are precisely $\lambda_i^{(1)}-(\lambda_i^{(1)})^2$. The intrinsic dimension of $\tilde{V}$ satisfies $d:=\tr(\tilde{V})/\|\tilde{V}\|_{2}\lesssim{r}$. Then, using Lemma \ref{lem:epm} and Theorem 7.7.1 in \citep{tropp2015introduction}, we acquire the proof.
\end{proof}

Finally, according to the higher-order Davis-Kahan theorem, i.e., Proposition \ref{lem:taylor},
$$\hat{U}^{(1)}(\hat{U}^{(1)})^{\top}-UU^{\top} = UL^{(0)}U^{\top}+R^{(0)}.$$
By Lemma \ref{lem:epm} and Lemma \ref{lem:mc}, $\|R^{(0)}\|_F=O_p(K^{-1})$, so we only need to calculate $\|UL^{(0)}U^{\top}\|_F^2=\|L^{(0)}\|_F^2$. Moreover, since $L^{(0)}$ is the linear term, we have
\begin{equation*}
    \begin{aligned}
       \E\left(\|L^{(0)}\|_F^2\right)&=\E\left[\tr(L^{(0)}L^{(0)})\right]=\E\left[\tr[(\sum_{k=1}^{K}L_k^{(0)}/K)^2]\right] \\
       &= K^{-2}\sum_{k=1}^{K}\E\left[\tr(L_k^{(0)}L_k^{(0)})\right] = K^{-1} \E\left(\|L_k^{(0)}\|_F^2\right).
    \end{aligned}
\end{equation*}
The second line is due to independence between the centered $L_k^{(0)}$, where $L^{(0)}=\sum_{k=1}^{K}L_k^{(0)}/K$ with
$$(L_k^{(0)})_{ij}=(L_k^{(0)})_{ji} = \frac{u_i^{\top}(\tilde{U}_k\tilde{U}^{\top}_k-\E(\tilde{U}_k\tilde{U}^{\top}_k))u_j}{\lambda^{(1)}_i-\lambda^{(1)}_j},\quad i\leq r,\, j>r,$$
and $0$ otherwise. Since $u_i^{\top}\E(\tilde{U}_k\tilde{U}^{\top}_k))u_j = 0$, $i\leq r$, $j>r$, we have for $\tilde{U}_k\tilde{U}^{\top}_k=\sum_{m=1}^{r}\tilde{u}^k_m(\tilde{u}^k_m)^{\top}$ that
    \begin{equation*}
        \begin{aligned}
            \|L_k^{(0)}\|^2_F &=\sum_{i=1}^{r}\sum_{j=r+1}^{p}\sum_{m=1}^{r} \frac{u_i^{\top}\tilde{u}_m(\tilde{u}_m)^{\top}u_ju_j^{\top}\tilde{u}_m(\tilde{u}_m)^{\top}u_i}{(\lambda^{(1)}_i-\lambda^{(1)}_j)^2}\\
                &= \sum_{i=1}^{r}\sum_{m=1}^{r} \frac{u_i^{\top}\tilde{u}_m(\tilde{u}_m)^{\top}U_{\perp}U_{\perp}^{\top}\tilde{u}_m(\tilde{u}_m)^{\top}u_i}{(\lambda^{(1)}_i-\lambda^{(1)}_{r+1})^2}\\
                &= \sum_{i=1}^{r}\sum_{m=1}^{r} \frac{\tr\left(u_iu_i^{\top}\tilde{u}_m(\tilde{u}_m)^{\top}\right)\tr\left(U_{\perp}U_{\perp}^{\top}\tilde{u}_m(\tilde{u}_m)^{\top}\right)}{(\lambda^{(1)}_i-\lambda^{(1)}_{r+1})^2}.
        \end{aligned}
    \end{equation*}
    In the second line, we use the fact that $\lambda_j^{(1)}=\lambda_{r+1}^{(1)}$ for all $j\geq r+1$, according to Lemma \ref{lem:epm}. Finally, by Proposition \ref{prop:rmt-eigenvector}, we have
$$\tr\left(u_iu_i^{\top}\tilde{u}_i(\tilde{u}_i)^{\top}\right)\xrightarrow{\text {a.s.}} \lambda^{(0)}_i,$$
$$\tr\left(u_iu_i^{\top}\tilde{u}_m(\tilde{u}_m)^{\top}\right)\xrightarrow{\text {a.s.}} 0, \quad i\neq m.$$
As $\lambda^{(1)}_{r+1}\rightarrow 0$, we have directly that
$$ \E\left(\|L_k^{(0)}\|^2_F\right) \rightarrow \sum_{i=1}^{r} \frac{\E\left[\tr\left(U_{\perp}U_{\perp}^{\top}\tilde{u}_i(\tilde{u}_i)^{\top}\right)\right]}{\lambda^{(1)}_i}\rightarrow \sum_{i=1}^{r} \frac{1-\lambda^{(1)}_i}{\lambda^{(1)}_i}.$$
The proof is complete. \hfill $\qedsymbol$

\subsection{Proof of Theorem \ref{theorem:2rdpca}}

To some extent, Theorem \ref{theorem:2rdpca} could be viewed as a corollary of Theorem \ref{theorem:3}, though some of the conditions do not exactly match, and can be in principle proved using Proposition \ref{prop:sub}. However, we choose to present the proof in an alternative way: for $\tilde{\Sigma}^{\cP}:=\hat{\Sigma}^{\cP}-(\hat{\sigma}^{(1)})^2I_p$, and $\hat{\Sigma}^{(2)}:=\tilde{\Sigma}^{\cP}\hat{U}^{(1)}(\hat{U}^{(1)})^{\top}\tilde{\Sigma}^{\cP}$, we carefully control each term in the following matrix decomposition:
\begin{equation}\label{eq:main-decomp}
      \hat{\Sigma}^{(2)} = \underbrace{\tilde{\Sigma}^{\cP}\hat{U}^{\cP}(\hat{U}^{\cP})^{\top}\tilde{\Sigma}^{\cP}}_{\text{signal}} + \underbrace{\tilde{\Sigma}^{\cP}\hat{\cU}^{\cP}\hat{L}^{(1)} (\hat{\cU}^{\cP})^{\top}\tilde{\Sigma}^{\cP}}_{\text{tangent first-order}} + \underbrace{\tilde{\Sigma}^{\cP}\hat{R}^{(1)} \tilde{\Sigma}^{\cP}}_{\text{higher-order}},
\end{equation}
where $\hat{\cU}^{\cP}:=(\hat{U}^{\cP}, \hat{U}^{\cP}_{\perp})$, and $\Delta^{(1)}=\hat{U}^{(1)}(\hat{U}^{(1)})^{\top}-\hat{U}^{\cP}(\hat{U}^{\cP})^{\top}=\hat{\cU}^{\cP}\hat{L}^{(1)} (\hat{\cU}^{\cP})^{\top}+\hat{R}^{(1)}$. The decomposition comes from Proposition \ref{lem:taylor}, the higher-order Davis-Kahan theorem, where $\|\hat{\Delta}^{(1)} \|_F=O_p(K^{-1/2})$, $\|\hat{L}^{(1)}\|_F=O_p(K^{-1/2})$, and $\|\hat{R}^{(1)} \|_F=O_p(K^{-1})$.

Clearly, $\hat{U}^{(2)}$ consists of the leading $r$ eigenvectors of $\hat{\Sigma}^{(2)}$. The reason for choosing this path is two-fold. First, it provides a clearer picture of the eigenvalue shifted subspace iteration in a statistical sense, as it enables precise tracking of each error component. Second, the analysis here also leads to results concerning eigenvalue estimation, and Corollary \ref{coro:eigenvalues} directly follows. 

First, the signal term $\tilde{\Sigma}^{\cP}\hat{U}^{\cP}(\hat{U}^{\cP})^{\top}\tilde{\Sigma}^{\cP}$ is of rank $r$. Its leading $r$ eigenvectors are the same as those of $\hat{\Sigma}^{\cP}$. Its non-trivial eigenvalues are $(\hat{\lambda}_i^{\cP}-(\hat{\sigma}^{(1)})^2)^2$ for $i\in [r]$. We shall show later in Section \ref{sec:proof-coro} that $\hat{\sigma}_{(1)}^2 = 1+O_p(K^{-1/2})$. We first take this as a given fact.

Meanwhile, the second part $\hat{T}^{(1)}:=\tilde{\Sigma}^{\cP}\hat{\cU}^{\cP}\hat{L}^{(1)} (\hat{\cU}^{\cP})^{\top}\tilde{\Sigma}^{\cP}$ is almost tangent to the signal part as $K\rightarrow \infty$. We decompose this part into the sum of four terms
\begin{equation}\label{eq:decomp}
    \begin{aligned}
    \hat{T}^{(1)}&= \underbrace{\hat{U}^{\cP}(\hat{U}^{\cP})^{\top}\hat{T}^{(1)}\hat{U}^{\cP}(\hat{U}^{\cP})^{\top}}_{(a)}+\underbrace{\hat{U}_{\perp}^{\cP}(\hat{U}_{\perp}^{\cP})^{\top}\hat{T}^{(1)}\hat{U}_{\perp}^{\cP}(\hat{U}_{\perp}^{\cP})^{\top}}_{(b)}\\
       &\quad + \underbrace{\hat{U}^{\cP}(\hat{U}^{\cP})^{\top}\hat{T}^{(1)}\hat{U}_{\perp}^{\cP}(\hat{U}_{\perp}^{\cP})^{\top}}_{(c)} + \underbrace{\hat{U}_{\perp}^{\cP}(\hat{U}_{\perp}^{\cP})^{\top}\hat{T}^{(1)}\hat{U}^{\cP}(\hat{U}^{\cP})^{\top}}_{(d)}.
    \end{aligned}
\end{equation}

The first two terms of \eqref{eq:decomp} share the same eigenspaces with $\tilde{\Sigma}^{\cP}\hat{U}^{\cP}(\hat{U}^{\cP})^{\top}\tilde{\Sigma}^{\cP}$. Meanwhile, the magitude of these two terms is of order $K^{-1/2}$. Due to Weyl's inequality, the leading eigenspace of $\tilde{\Sigma}^{\cP}\hat{U}^{\cP}(\hat{U}^{\cP})^{\top}\tilde{\Sigma}^{\cP}+(a)+(b)$ would still be $\hat{U}^{\cP}$ with probability tending to 1.

It suffices to control the ``noise terms", i.e., (c), (d) and $\tilde{\Sigma}^{\cP}\hat{R}^{(1)} \tilde{\Sigma}^{\cP}$. We control (c) in the following Lemma \ref{lem:cross}, while (d) can be similarly handled.

\begin{lemma}\label{lem:cross}
Under Assumption \ref{assump:1}, for $\tilde{\Sigma}^{\cP}:=\hat{\Sigma}^{\cP}-(\hat{\sigma}^{(1)})^2I_p$, we have
$$\left\|\underbrace{\hat{U}^{\cP}(\hat{U}^{\cP})^{\top}\tilde{\Sigma}^{\cP}\hat{\cU}^{\cP}\hat{L}^{(1)} (\hat{\cU}^{\cP})^{\top}\tilde{\Sigma}^{\cP}\hat{U}_{\perp}^{\cP}(\hat{U}_{\perp}^{\cP})^{\top}}_{\text{cross-term}}\right\|_{F} = O_p\left(K^{-1}\right).$$
\end{lemma}

\begin{proof}
Clearly, the eigenvectors of $\tilde{\Sigma}^{\cP}$ are the same as those of $\hat{\Sigma}^{\cP}$. So we can write $\tilde{\Sigma}^{\cP} = \hat{\cU}^{\cP} \tilde{\Lambda}^{\cP}(\hat{\cU}^{\cP})^{\top}$, where $\tilde{\Lambda}^{\cP} = \diag(\tilde{\lambda}^{\cP}_{1},\dots \tilde{\lambda}^{\cP}_{p})$ with $\tilde{\lambda}^{\cP}_{i}:=\hat{\lambda}^{\cP}_{i}-(\hat{\sigma}^{(1)})^2$. Then the cross-term equals
\begin{equation*}
\tilde{C}^{(1)}=\underbrace{\hat{U}^{\cP}(\hat{U}^{\cP})^{\top} \hat{\cU}^{\cP}}_{(\hat{U}^{\cP},0)} \tilde{\Lambda}^{\cP} \hat{L}^{(1)}\tilde{\Lambda}^{\cP}\underbrace{(\hat{\cU}^{\cP})^{\top}\hat{U}_{\perp}^{\cP}(\hat{U}_{\perp}^{\cP})^{\top}}_{(0,\hat{U}_{\perp}^{\cP})^{\top}}.
\end{equation*}

For $$\hat{L}^{(1)}:= \left(\begin{array}{cc}0_{r\times r} & (\hat{K}^{(1)})^{\top} \\ \hat{K}^{(1)} & 0_{(p-r)\times (p-r)}\end{array}\right), \quad \hat{\Lambda}^{\cP}:= \left(\begin{array}{cc}\tilde{\Lambda}^{\cP}_1 & 0_{r\times (p-r)} \\ 0_{(p-r)\times r} & \tilde{\Lambda}^{\cP}_2\end{array}\right),$$
we have
\begin{equation*}
\begin{aligned}
\tilde{C}^{(1)} &=  \left(\hat{U}^{\cP},0_{p\times (p-r)}\right)  \left(\begin{array}{cc}0_{r\times r} & \tilde{\Lambda}^{\cP}_1(\hat{K}^{(1)})^{\top}\tilde{\Lambda}^{\cP}_2 \\\tilde{\Lambda}^{\cP}_2 \hat{K}^{(1)}\tilde{\Lambda}^{\cP}_1 & 0_{(p-r)\times (p-r)}\end{array}\right) \left(\begin{array}{c}0_{(p-r)\times p}  \\ (\hat{U}_{\perp}^{\cP})^{\top}\end{array}\right)\\
&=\left(\begin{array}{cc}0_{r\times r} & \hat{U}^{\cP}\tilde{\Lambda}^{\cP}_1(\hat{K}^{(1)})^{\top}\tilde{\Lambda}^{\cP}_2 (\hat{U}_{\perp}^{\cP})^{\top}\\0_{(p-r)\times r} & 0_{(p-r)\times (p-r)}\end{array}\right).
\end{aligned}
\end{equation*}
Hence, $\|\tilde{C}^{(1)}\|_F=\|\hat{U}^{\cP}\tilde{\Lambda}^{\cP}_1(\hat{K}^{(1)})^{\top}\tilde{\Lambda}^{\cP}_2 (\hat{U}_{\perp}^{\cP})^{\top}\|_F$. We only need to bound both $\|\tilde{\Lambda}^{\cP}_1\|_{2}$ and $\|\tilde{\Lambda}^{\cP}_2\|_{2}$. According to Lemma 3 in \cite{fan2019distributed}, i.e., non-asymptotic Bai-Yin Theorem, we know that $\|\hat{\Sigma}^{\cP}-\Sigma\|_{2} = O_p(\sqrt{p/Kn})=O_p(K^{-1/2})$. Hence, $\|\tilde{\Lambda}^{\cP}_1\|_{2}=O_p(1)$ and $\|\tilde{\Lambda}^{\cP}_2\|_{2}=O_p(K^{-1/2})$ are obtained. Finally, based on Lemma \ref{lem:F2} and the fact that $\|\hat{K}^{(1)}\|_F\asymp \|\hat{L}^{(1)}\|_F=O_p(K^{-1/2})$, Lemma \ref{lem:cross} is proved. 
\end{proof}

Finally, the Frobenius norm of the higher-order term $\tilde{\Sigma}^{\cP}\hat{R}^{(1)} \tilde{\Sigma}^{\cP}$ tends to $0$ at the rate of $K^{-1}$. With these results in mind, using the Davis-Kahan Theorem \citep{yu2015useful}, we know that $\|\hat{U}^{(2)}(\hat{U}^{(2)})^{\top}-\hat{U}^{\cP}(\hat{U}^{\cP})^{\top}\|^2_F=O_p(K^{-2})$. The proof is complete.  \hfill $\qedsymbol$

\subsection{Proof of Corollary \ref{coro:eigenvalues}}\label{sec:proof-coro}
It is a direct corollary of the matrix perturbation in the proof of Theorem \ref{theorem:2rdpca}. Notice that the non-trivial eigenvalues of the signal part in \eqref{eq:main-decomp} are exactly $(\hat{\lambda}_i^{\cP}-(\hat{\sigma}^{(1)})^2)^2$ for $i\in [r]$, where $\hat{\lambda}_i^{\cP}$ has been extensively studied; one can refer to, e.g., Theorem 2.13 in \cite{couillet2022random}. 

We now prove that $\hat{\sigma}_{(1)}^2 = 1+O_p(K^{-1/2})$. Since $\hat{U}_{\perp}^{(1)}(\hat{U}_{\perp}^{(1)})^{\top}=I_p-\hat{U}^{(1)}(\hat{U}^{(1)})^{\top}$ and $\hat{U}_{\perp}^{\cP}(\hat{U}_{\perp}^{\cP})^{\top}=I_p - \hat{U}^{\cP}(\hat{U}^{\cP})^{\top}$, we have
$$\left\|\hat{U}_{\perp}^{(1)}(\hat{U}_{\perp}^{(1)})^{\top}-\hat{U}_{\perp}^{\cP}(\hat{U}_{\perp}^{\cP})^{\top}\right\|_F=\left\|\hat{U}^{(1)}(\hat{U}^{(1)})^{\top}-\hat{U}^{\cP}(\hat{U}^{\cP})^{\top}\right\|_F=O_p(K^{-1/2}).$$
Hence, for $\tr(\hat{\Sigma}^{\cP}\hat{U}_{\perp}^{(1)}(\hat{U}_{\perp}^{(1)})^{\top})=\tr(\hat{\Sigma}^{\cP}\hat{U}_{\perp}^{\cP}(\hat{U}_{\perp}^{\cP})^{\top})+\tr[\hat{\Sigma}^{\cP}(\hat{U}_{\perp}^{(1)}(\hat{U}_{\perp}^{(1)})^{\top}-\hat{U}_{\perp}^{\cP}(\hat{U}_{\perp}^{\cP})^{\top})]$,
we know that
$$\frac{1}{p-r}\left|\tr\left[\hat{\Sigma}^{\cP}\left(\hat{U}_{\perp}^{(1)}(\hat{U}_{\perp}^{(1)})^{\top}-\hat{U}_{\perp}^{\cP}(\hat{U}_{\perp}^{\cP})^{\top}\right)\right]\right|\leq \frac{\|\hat{\Sigma}^{\cP}\|_F}{p-r}\left\|\hat{U}_{\perp}^{(1)}(\hat{U}_{\perp}^{(1)})^{\top}-\hat{U}_{\perp}^{\cP}(\hat{U}_{\perp}^{\cP})^{\top}\right\|_F=O_p(K^{-1/2}).$$

We only need to show that $\tr(\hat{\Sigma}^{\cP}\hat{U}_{\perp}^{\cP}(\hat{U}_{\perp}^{\cP})^{\top})/(p-r) = \sum_{j=r+1}^{p} \hat{\lambda}^{\cP}_j/(p-r)\rightarrow 1$. Using Lemma 3 in \cite{fan2019distributed}, we obtain the fact that 
$$\left|\frac{1}{p-r}\sum_{j=r+1}^{p} \hat{\lambda}^{\cP}_j-1\right| \leq \frac{1}{p-r}\sum_{j=r+1}^{p} \left|\hat{\lambda}^{\cP}_j-1\right| \leq \max_{j\geq r+1} \left|\hat{\lambda}^{\cP}_j-1\right| = O_p(K^{-1/2}).$$

In the end, using Lemma 3 in \cite{fan2019distributed}, we know that $\hat{\lambda}_i^{\cP}-l_i-1 = O_p(K^{-1/2})$. Meanwhile, using the Delta method, we know from the proof of Theorem \ref{theorem:2rdpca} that $\hat{\lambda}_i^{\cP}-\hat{l}^{(2)}_i-1=O_p(K^{-1/2})$. The proof is complete. \hfill $\qedsymbol$

\subsection{Proof of Corollary \ref{coro:an}}
Corollary \ref{coro:an} comes from combining both Theorem \ref{theorem:2rdpca} and Proposition \ref{lem:lt}, i.e., Lemma S.7 from \cite{li2024tpca}. Using Proposition \ref{lem:lt}, we first separate $\hat{U}^{\cP}(\hat{U}^{\cP})^{\top}-UU^{\top}$ into two terms. We denote the linear term as $\hat{L}^{\cP}$ and the higher-order term as $\hat{R}^{\cP}$. The bilinear form of the linear term has asymptotic variance given in \eqref{eq:a_variance}.

As we can see in the second term of \eqref{eq:a_variance}, the asymptotic variance of the bilinear form is related to the fourth moment of the underlying distribution. Fortunately, under the Gaussian distribution, $\nu_4=3$ and the second term equals zero. It suffices to calculate the first term, i.e.,
\begin{equation}\label{eq:rij1}
    \sigma_{(u,v)}^2:=\sum_{i=1}^{r}\frac{(l_i+1)}{l_i^2}\sum_{j=r+1}^{p}\rho^2_{ij},
\end{equation}
where $\rho_{ij}:=(U^{\top}u)_i(U^{\top}v)_j+(U^{\top}u)_j(U^{\top}v)_i$, and the subscript $i$ means the $i$-th element of the vector. Notice that
\begin{equation}\label{eq:rij2}
    \begin{aligned}
        &\sum_{j=r+1}^{p}\rho^2_{ij}=\sum_{j=r+1}^{p} (U^{\top}u)^2_i(U^{\top}v)^2_j + \sum_{j=r+1}^{p}(U^{\top}u)^2_j(U^{\top}v)^2_i + 2\sum_{j=r+1}^{p} (U^{\top}u)_i(U^{\top}v)_i(U^{\top}u)_j(U^{\top}v)_j\\
        & = (U^{\top}u)^2_i v^{\top}(I_p-UU^{\top})v + (U^{\top}v)^2_i u^{\top}(I_p-UU^{\top})u + 2(U^{\top}u)_i(U^{\top}v)_i  u^{\top}(I_p-UU^{\top})v.
        \end{aligned}
\end{equation}
Combining both \eqref{eq:rij1} and \eqref{eq:rij2}, we have
\begin{equation}\label{eq:sigma2}
    \begin{aligned}
    \sigma_{(u,v)}^2&= u^{\top}[I_p-UU^{\top}]u\cdot\sum_{i=1}^{r}(l_i+1)(u_i^{\top}v)^2/l_i^2 + v^{\top}[I_p-UU^{\top}]v\cdot\sum_{i=1}^{r}(l_i+1)(u_i^{\top}u)^2/l_i^2 \\
    &\quad + 2u^{\top}[I_p-UU^{\top}]v\cdot\sum_{i=1}^{r}(l_i+1)(u_i^{\top}u)(u_i^{\top}v)/l_i^2.
    \end{aligned}
\end{equation}

Meanwhile, the higher-order term $\hat{R}^{\cP}$ is of order $p/Kn$, in terms of the Frobenius norm. Following the proof of Theorem \ref{theorem:2rdpca}, we know that $\hat{U}^{(2)}(\hat{U}^{(2)})^{\top}$ converges to $\hat{U}^{\cP}(\hat{U}^{\cP})^{\top}$ at a rate of $p/Kn$. In the end, from the proof of Theorem \ref{theorem:2rdpca} and Corollary \ref{coro:eigenvalues}, we have $(\hat{\sigma}_{(u,v)}^{(T)})^2-\sigma_{(u,v)}^2 = o(1)$ for
   \begin{equation}\label{eq:hatsigma2} 
    \begin{aligned}
            (\hat{\sigma}_{(u,v)}^{(t)})^2: &= u^{\top}[I_p-\hat{U}^{(t)}(\hat{U}^{(t)})^{\top}]u\cdot\sum_{i=1}^{r}[\hat{l}^{(t)}_i+(\hat{\sigma}^{(t-1)})^2][(\hat{u}^{(t)}_i)^{\top}v]^2/(\hat{l}_i^{(t)})^2\\
            &\quad + v^{\top}[I_p-\hat{U}^{(t)}(\hat{U}^{(t)})^{\top}]v\cdot\sum_{i=1}^{r}[\hat{l}^{(t)}_i+(\hat{\sigma}^{(t-1)})^2][(\hat{u}^{(t)}_i)^{\top}u]^2/(\hat{l}_i^{(t)})^2 \\
    &\quad + 2u^{\top}[I_p-\hat{U}^{(t)}(\hat{U}^{(t)})^{\top}]v\cdot\sum_{i=1}^{r}[\hat{l}^{(t)}_i+(\hat{\sigma}^{(t-1)})^2][(\hat{u}^{(t)}_i)^{\top}u][(\hat{u}^{(t)}_i)^{\top}v]/(\hat{l}_i^{(t)})^2.
    \end{aligned}
\end{equation}
The proof is complete using Slutsky's Theorem. \hfill $\qedsymbol$

\subsection{Proof of Theorem \ref{theorem:3}}
   First, we discuss the case 2 when $\delta \geq c_2>0$. Due to the potentially infinite iteration steps, we need to give a uniform bound for the ratios $|(\hat{\lambda}_{r+1}^{\cP}-(\hat{\sigma}^{(t)})^2)/(\hat{\lambda}_{r}^{\cP}-(\hat{\sigma}^{(t)})^2)|$ and $|((\hat{\sigma}^{(t)})^2-\hat{\lambda}_{p}^{\cP})/(\hat{\lambda}_{r}^{\cP}-(\hat{\sigma}^{(t)})^2)|$ over all $t\geq1$. We only discuss $|(\hat{\lambda}_{r+1}^{\cP}-(\hat{\sigma}^{(t)})^2)/(\hat{\lambda}_{r}^{\cP}-(\hat{\sigma}^{(t)})^2)|$ here, as the other one can be handled similarly.

    For both the numerator $\hat{\lambda}_{r+1}^{\cP}-(\hat{\sigma}^{(t)})^2$ and denominator $\hat{\lambda}_{r}^{\cP}-(\hat{\sigma}^{(t)})^2$, the random error lies in both the eigenvalue estimation, i.e., $|\hat{\lambda}_{i}^{\cP}-\lambda_i|$, and the eigenvalue shifting, i.e., $|(\hat{\sigma}^{(t)})^{2}-\sigma^2|$. For the error in eigenvalue estimation, we know from Lemma 3 in \cite{fan2019distributed} that $\max_{i} |\hat{\lambda}_{i}^{\cP}-\lambda_i|=O_p(\sqrt{pr/Kn})$, which is free of $t$. Then, we control the statistical error in eigenvalue shifting. We know that 
    $$(\hat{\sigma}^{(t)})^{2}-\sigma^2 = \left[(\hat{\sigma}^{(t)})^{2}-\tr(\hat{\Sigma}^{\cP})/p\right]+\left[\tr(\hat{\Sigma}^{\cP})/p-\tr(\Sigma)/p\right]+\left[\tr(\Sigma)/p-\sigma^2\right].$$
    Due to the boundedness of eigenvalues in Assumption \ref{assump:2}, we have $(\hat{\sigma}^{(t)})^{2}-\tr(\hat{\Sigma}^{\cP})/p \lesssim \|\hat{\Sigma}^{\cP}\|_{2}/p = O_p(p^{-1})$. Similarly, $\tr(\Sigma)/p-\sigma^2=O(p^{-1})$. As a result, $[(\hat{\sigma}^{(t)})^{2}-\tr(\hat{\Sigma}^{\cP})/p]+[\tr(\Sigma)/p-\sigma^2]$ is uniformly controlled by $\sqrt{pr/Kn}$ for all $t\geq 1$, given the assumption that $Kn\leq C_2 p^3$.

    Meanwhile, note that $\tr(\hat{\Sigma}^{\cP}) = \sum_{k=1}^{K}\sum_{i=1}^{n}(x_i^{k})^{\top}x_i^{k}/(Kn)$ is also free of $t$. Under Assumption \ref{assump:2}, according to the Hason-Wright inequality, e.g., Theorem 6.2.3 in \cite{vershynin2018high}, $(x_i^{k})^{\top}x_i^{k}/p-\tr(\Sigma)/p =O_p(1)$. Therefore, the statistical error in eigenvalue shifting could be absorbed into the eigenvalue estimation error, which is of order $O_p(\sqrt{pr/Kn})$. In the cases where $\delta \geq c_2>0$ and $\lambda_{r}\geq \sigma^2+\delta+c_1$, as $Kn\gg pr$ so that both the eigenvalue estimation and eigenvalue shifting errors tend to $0$, we have with probability tending to $1$ that
$$\left|\frac{\hat{\lambda}_{r+1}^{\cP}-(\hat{\sigma}^{(t)})^2}{\hat{\lambda}_{r}^{\cP}-(\hat{\sigma}^{(t)})^2}\right|\leq c_3<1, \quad\text{for all}\quad t\geq1.$$

Then, using Proposition \ref{prop:sub} recursively for $t=1,\dots, T$, we eventually have $\|U^{(T)}(U^{(T)})^{\top}-U^{\cP}(U^{\cP})^{\top}\|_2 =O_p(c_3^{T-1})$. For $U^{(T)}$ to be statistically efficient, we need $c_3^{T}\ll \sqrt{pr/Kn}$, that is, $T\gg \log(Kn/pr)$. Then, the difference between $U^{(T)}$ and $U^{\cP}$ is much smaller than the estimation error of $U^{\cP}$, which is $O_p(\sqrt{pr/Kn})$, and the case 2 is proved.

As for the case 1, if $\alpha< 1/2$, then $\delta$ is still the main term in the numerator of the shifted eigenvalue ratio, which is analogous to the case $2$. In this case, we need $\delta^{T-1}\ll \sqrt{pr/Kn}$, which means $T-1>1/2\alpha$ and hence $T>(2\alpha+1)/2\alpha$. On the other hand, if $\alpha\geq 1/2$, then the eigenvalue estimation error is the main term, which is also of order $\sqrt{pr/Kn}$. In this case, we need two more steps to alleviate the potential bias from $\hat{U}^{(1)}$, resulting in $T=3$. As a result, in the case 1, given $T=\max(3,\lceil (2\alpha+1)/2\alpha +\varepsilon\rceil)$ for some $\varepsilon>0$, we have $\lim\limits_{k\rightarrow \infty} \lim\limits_{n,p\rightarrow \infty} \mathcal{V}\left(\hat{U}^{(T)}\right) =1$. The proof is complete. \hfill $\qedsymbol$

\section{Proof of Corollary \ref{theorem:ellip}}

The proof is identical to the proof of Theorem \ref{theorem:3}, only here we work with the Kendall's $\tau$ matrix, instead of the covariance matrix. Define the population Kendall's $\tau$ matrix as
$$\Sigma^{\tau}=\E_{i\neq j}\left[\frac{(x_{i}^{k}-x_{j}^{k})(x_{i}^{k}-x_{j}^{k})^{\top}}{\|x_{i}^{k}-x_{j}^{k}\|^2}\right].$$
By Proposition 2.1 in \cite{han2018eca}, the leading $r$ eigenvectors of $\Sigma$ and $\Sigma^{\tau}$ are identical, up to some orthogonal transformation, with the same descending order of the corresponding eigenvalues under mild conditions. In fact, we have
\begin{equation}\label{eq:eg-kendall}    \lambda_j^{\tau}:=\lambda_j(\Sigma^{\tau})=\E\left(\frac{\lambda_j(\Sigma)Y_j^2}{\sum_{i=1}^{p}\lambda_i(\Sigma)Y_i^2}\right),
\end{equation}
where $(Y_1,\dots,Y_{p})^{\top}\sim N_{p}(0,I_{p})$ is a standard Gaussian random vector. That is to say, estimating the leading eigenspace of $\Sigma^{\tau}$ is equivalent to estimating that of the scatter matrix $\Sigma$.

\begin{proposition}[Lemma A.2 and A.5 from \cite{he2022distributed}]\label{lem:ext}
    Under Assumption \ref{assum:ellip}, we have
    \begin{equation}\label{eq:ecaindierror}
    \left\|\left\|\tilde{\Sigma}^{\tau}_{k}-\Sigma^{\tau}\right\|_2\right\|_{\psi_1}\lesssim
    \sqrt{\frac{\log p}{n}},
    \end{equation}
    where $\Sigma^{\tau}$ and $\tilde{\Sigma}^{\tau}_{k}$ stand for the population and sample versions of the local Kendall's $\tau$ matrix. Moreover, recall that $\Sigma$ is the scatter matrix. For the $r$-th eigengap of $\Sigma^{\tau}$, we have the following bound when $p$ is sufficiently large:
    \begin{equation}\label{eq:ecagapbound}
       \lambda_{r}(\Sigma^{\tau})- \lambda_{r+1}(\Sigma^{\tau})\gtrsim\frac{\lambda_{r}(\Sigma)- \lambda_{r+1}(\Sigma)}{\tr(\Sigma)}\gtrsim p^{-1}.
    \end{equation}
\end{proposition}

We can see by definition that $\tr(\Sigma^{\tau})=1$. Indeed, \eqref{eq:eg-kendall} essentially states that Kendall's $\tau$ matrix rescales the eigenvalues of the scatter matrix $\Sigma$ approximately by $p^{-1}$. Hence, \eqref{eq:ecagapbound} ensures a positive eigenvalue gap after the rescaling.

The proof is then stated as follows. Define the average matrix of the local sample Kendall's $\tau$ estimators as $\tilde{\Sigma}^{\tau}:=\sum_{k=1}^{K}\tilde{\Sigma}^{\tau}_{k}/K$. By combining Proposition \ref{lem:subexp} and \eqref{lem:ext}, we know that
$$  \left\|\left\|\tilde{\Sigma}^{\tau}-\Sigma^{\tau}\right\|_2\right\|_{\psi_1}\lesssim
    \sqrt{\frac{\log p}{Kn}}.$$
By the Davis-Kahan Theorem and \eqref{eq:ecagapbound}, we know that 
$$ \left\|\tilde{U}_{\tau}\hat{U}_{\tau}^{\top}-UU^{\top}\right\|_F=O_p\left(\sqrt{\frac{p\log p r}{Kn}}\right),$$
where $\tilde{U}_{\tau}$ consists of the leading $r$ eigenvectors of $\tilde{\Sigma}^{\tau}$. Similar to the proof of Theorem \ref{theorem:3}, it suffices to control the error between $\hat{U}^{(t)}_{\tau}\left(\hat{U}_{\tau}^{(t)}\right)^{\top}$ and $\tilde{U}_{\tau}\hat{U}_{\tau}^{\top}$ using Proposition \ref{prop:sub}. As we assume homogeneous noise eigenvalues in Assumption \ref{assum:ellip}, the proof is then identical to $\alpha\geq 1/2$ in case 1 of Theorem \ref{theorem:3}, i.e., the eigenvalue estimation error is the main term in the numerator of the shifted eigenvalue ratio, and $T=3$ is sufficient for convergence under Assumption \ref{assum:ellip}. The proof is complete. \hfill $\qedsymbol$

\section{Some Useful Results}\label{sec:sur}

For convenience, we state some useful results in this section, which are used throughout the theoretical analysis.

\begin{lemma}\label{lem:F2}
    For two real matrices $A$ and $B$, we have $\|AB\|_F\leq \|A\|_{2}\|B\|_F$.
\end{lemma}
\begin{proof}
    Let $B:=(b_1,\dots,b_n)$, we have
    \begin{equation*}
        \begin{aligned}
            \|AB\|_F^2&=\sum_{j=1}^{n}\|Ab_j\|^2_2\leq
            \sum_{j=1}^{n}\|A\|^2_{2}\|b_j\|^2_2\\
            &=\|A\|^2_{2} \left(\sum_{j=1}^{n}\|b_j\|^2_2\right)=\|A\|^2_{2}\|B\|^2_F.
        \end{aligned}
    \end{equation*}
\end{proof}

\begin{proposition}[Theorem 2.5 from \cite{bosq2000stochastic}]\label{lem:subexp}
	For independent random vectors $\{X_i\}_{i=1}^{n}$ in a separable Hilbert space with norm $\|\cdot\|$, if $\E(X_i) = 0$ and $\|\|X_i\|\|_{\psi_1}\leq L_i<\infty$, we have
    $$\left\|\left\|\sum_{i=1}^{n}X_i\right\|\right\|_{\psi_1}\lesssim \left(\sum_{i=1}^{n}L_i^2\right)^{1/2}.$$
\end{proposition}

\begin{proposition}[Higher-order Davis-Kahan theorem, from Lemma 2 of \cite{fan2019distributed}]\label{lem:taylor}
Let $\Sigma$ and $\hat{\Sigma}$ be $p\times p$ symmetric matrices with non-increasing eigenvalues $\lambda_1\geq \dots\geq \lambda_p$ and $\hat{\lambda}_1\geq \dots\geq \hat{\lambda}_p$, respectively. Let $\{u_i\}_{i=1}^{p}$, $\{\hat{u}_i\}_{i=1}^{p}$ be the corresponding eigenvectors such that $\Sigma u_i = \lambda_iu_i$ and $\hat{\Sigma} u_i = \hat{\lambda}_i\hat{u}_i$. Fix $s\in \{0,\dots,p-r\}$, let $\delta=\min(\lambda_s-\lambda_{s+1},\lambda_{s+r}-\lambda_{s+r+1})>0$, with $\lambda_0=+\infty$ and $\lambda_{p+1}=-\infty$. For $E = \hat{\Sigma}-\Sigma$, we denote $S=\{s+1,\dots,s+r\}$ and $S^c=\{1,\dots,p\}\setminus S$, and define $L$ of shape $p\times p$, whose entries are
$$L_{ij}=L_{ji}=
\left\{
    \begin{array}{lc}
        (u_i^{\top} E u_j)/(\lambda_i-\lambda_j), & i\in S,\, j\in S^c, \\
       0,& \text{otherwise}.\\
    \end{array}
\right.$$
Then, let $U = (u_1,\dots, u_p)$ and $\hat{U} = (\hat{u}_1,\dots, \hat{u}_p)$, while $U_S = (u_{s+1},\dots, u_{s+r})$ and $\hat{U}_S = (\hat{u}_{s+1},\dots, \hat{u}_{s+r})$. We claim that
$$\hat{U}_S\hat{U}_S^{\top}-U_SU_S^{\top} = ULU^{\top}+R,$$
where $\|L\|_F\leq (2r)^{1/2}\|E\|_{2}/\delta$. In addition, when $\|E\|_{2}/\delta\leq 1/10$, $\|R\|_F\leq 24r^{1/2}(\|E\|_{2}/\delta)^2$.

\end{proposition}
\begin{proof}
	Most claims in Proposition \ref{lem:taylor} follow directly from Lemma 2 in \cite{fan2019distributed}. We only need to check that $\|L\|_F\leq (2r)^{1/2}\|E\|_{2}/\delta$. To see this, note that
	$$\|L\|_F\leq  \frac{\sqrt{2}}{\delta}\left(\sum_{i=1}^{r}\left\|(U^{\top}EU)_{i,.}\right\|^2\right)^{1/2},$$
	where $A_{i,.}$ means the $i$-th row of matrix $A$ and $\|\cdot\|$ is the vector $l_2$ norm. Then, since $\|(U^{\top}EU)_{i,.}\|\leq \|U^{\top}EU\|_{2} = \|E\|_{2}$, we obtain the proof.
\end{proof}

\begin{proposition}[Lemma S.7 from \cite{li2024tpca}]\label{lem:lt}
    Under the settings of Proposition \ref{lem:taylor}, if we further set $\hat{\Sigma}=\sum_{k=1}^{n}Az_kz_k^{\top}A^{\top}/n$ where $\Sigma=AA^{\top}$, while the mutually independent random vectors $\{z_k\}_{k=1}^{n}$ consist of $p$ i.i.d. random variables $\{z_{k,m}\}_{m=1}^{p}$, such that $\E(z_{k,m})=0$, $\E(z^2_{k,m})=1$ and $\E(z^4_{k,m})=\nu_4$, let $E=\hat{\Sigma}-\Sigma$, take $u$, $v\in S^{p-1}$ and we have,
   $$\left\langle u, (\hat{U}_S\hat{U}_S^{\top}-U_SU_S^{\top})v\right\rangle =\frac{1}{n}\sum_{k=1}^{n}L_{S,k} +R_S,$$
    where $\{L_{S,k}\}_{k=1}^{n}$ are i.i.d. random variables with $\E(L_{S,k})=0$. Set $\rho_{ij}:=(U^{\top}u)_i(U^{\top}v)_j+(U^{\top}u)_j(U^{\top}v)_i$, and the subscript $i$ means the $i$-th element of the vector. Let $\gamma_i := A^{\top}u_i$ and $\gamma_{im}:=(A^{\top}u_i)_m$, we have
    \begin{equation}\label{eq:a_variance}
        \E(L^2_{S,k})=\sigma_S^2:=\sum_{i\in S}\sum_{j\in S^c}\frac{\rho^2_{ij}\lambda_i\lambda_j}{(\lambda_i-\lambda_j)^2}+(\nu_4-3)\sum_{m=1}^{p}\left( \sum_{i\in S}\sum_{j\in S^c} \frac{\rho_{ij}\gamma_{im} \gamma_{jm}}{\lambda_{i}-\lambda_{j}}\right)^2.
    \end{equation}
    Meanwhile, $R_S\lesssim r^{1/2}(\|E\|_2/\delta)^2$ within the event $\{\|E\|_2/\delta\leq 1/10\}$. 
\end{proposition}

\begin{lemma}[Subspace error]\label{lemma:subspace-error}
For two projection matrices $\hat{U}\hat{U}^{\top}$ and $UU^{\top}$, where $\hat{U}$ and $U$ are $p\times r$ column orthogonal matrices, we have
$$\frac{1}{2}\left\|\hat{U}\hat{U}^{\top}-UU^{\top}\right\|^2_F= r-\tr\left(\hat{U}\hat{U}^{\top}UU^{\top}\right).$$
\end{lemma}

\begin{proposition}[Subspace iteration, from Theorem 8.2.2 of \cite{golub2013matrix}]\label{prop:sub}
Let $A$ be a $p\times p$ real symmetric matrix with eigenvalues $|\lambda_1|\geq\cdots \geq |\lambda_r|>|\lambda_{r+1}|\geq  \cdots \geq|\lambda_p|$, and $U_r\in \reals^{p\times r}$ consists of eigenvectors corresponding to $|\lambda_1|\geq\cdots \geq |\lambda_r|$. Assume that $\|U_0U_0^{\top}-U_rU_r^{\top}\|_2<1$. Then, after one subspace iteration step:
$$Z_1 = AU_0,\quad U_1R_1 = Z_1 \quad\text{(QR factorization)},$$
we have
$$\left\|U_1U_1^{\top}-U_rU_r^{\top}\right\|_2\leq \left|\frac{\lambda_{r+1}}{\lambda_r}\right|\frac{\|U_0U_0^{\top}-U_rU_r^{\top}\|_2}{(1-\|U_0U_0^{\top}-U_rU_r^{\top}\|_2^2)^{1/2}}.$$

\end{proposition}

\begin{proposition}[Eigenvalue estimation, from Theorem 2.13 of \cite{couillet2022random}]\label{prop:rmt-eigenvalue}
    Under the spiked covariance model in this work with $p/n\rightarrow c$, let $\hat{\lambda}_i$ be the $i$-th largest eigenvalue of $\hat{\Sigma}$. We have
    $$\hat{\lambda}_i \xrightarrow{\text {a.s.}} \begin{cases}\rho_i=1+l_i+c (1+l_i)/l_i >(1+\sqrt{c})^2, &  l_i>\sqrt{c}, \\ (1+\sqrt{c})^2, &  l_i \leq \sqrt{c}.\end{cases}$$
\end{proposition}

\begin{proposition}[Eigenvector alignment, from Theorem 2.14 of \cite{couillet2022random}]\label{prop:rmt-eigenvector}
    Under the spiked covariance model in this work with $p/n\rightarrow c$, let $U=(u_1,\dots,u_r)$ be the first $r$ eigenvectors of $\Sigma$, and $\hat{U}=(\hat{u}_1,\dots,\hat{u}_r)$ the first $r$ eigenvectors of $\hat{\Sigma}$. Assume that $l_1>\dots>l_r>0$ are all distinct. Then, for unit norm deterministic vectors $a$, $b\in \reals^p$, we have
   $$a^{\top} \hat{u}_i \hat{u}_i^{\top} b-a^{\top} u_i u_i^{\top} b \frac{1-c l_i^{-2}}{1+c l_i^{-1}} 1_{l_i>\sqrt{c}} \xrightarrow{\text {a.s.}} 0.$$
\end{proposition}

\section{Additional Numerical Results}\label{app:num}

In this section, we report additional numerical results and details. 

\begin{table}
\centering
\caption{Basic dimensional information on the remaining benchmark datasets. We report $(N,p)$, where $N$ is the total sample size and $p$ is the feature dimension.}
\label{tab:Np}
\begin{tabular}{|c|c|c|c|c|}
\hline
  road-safety & MiniBooNE & jannis & Ailerons & yprop\_4\_1 \\\hline
  (111762, 32)&(72998, 50)&(57580, 54)&(13750, 33)&(8885, 42)\\
\hline
 Bioresponse & Allstate & Mercedes\_Benz & topo\_2\_1 & superconduct \\\hline
(3434, 419)&(188318, 124)&(4209, 359)&(8885, 255)&(21263, 79)\\
\hline
\end{tabular}
\end{table}

\begin{figure}
	\centering

    \begin{minipage}{0.9\linewidth}
		\centering
	\includegraphics[width=0.9\linewidth]{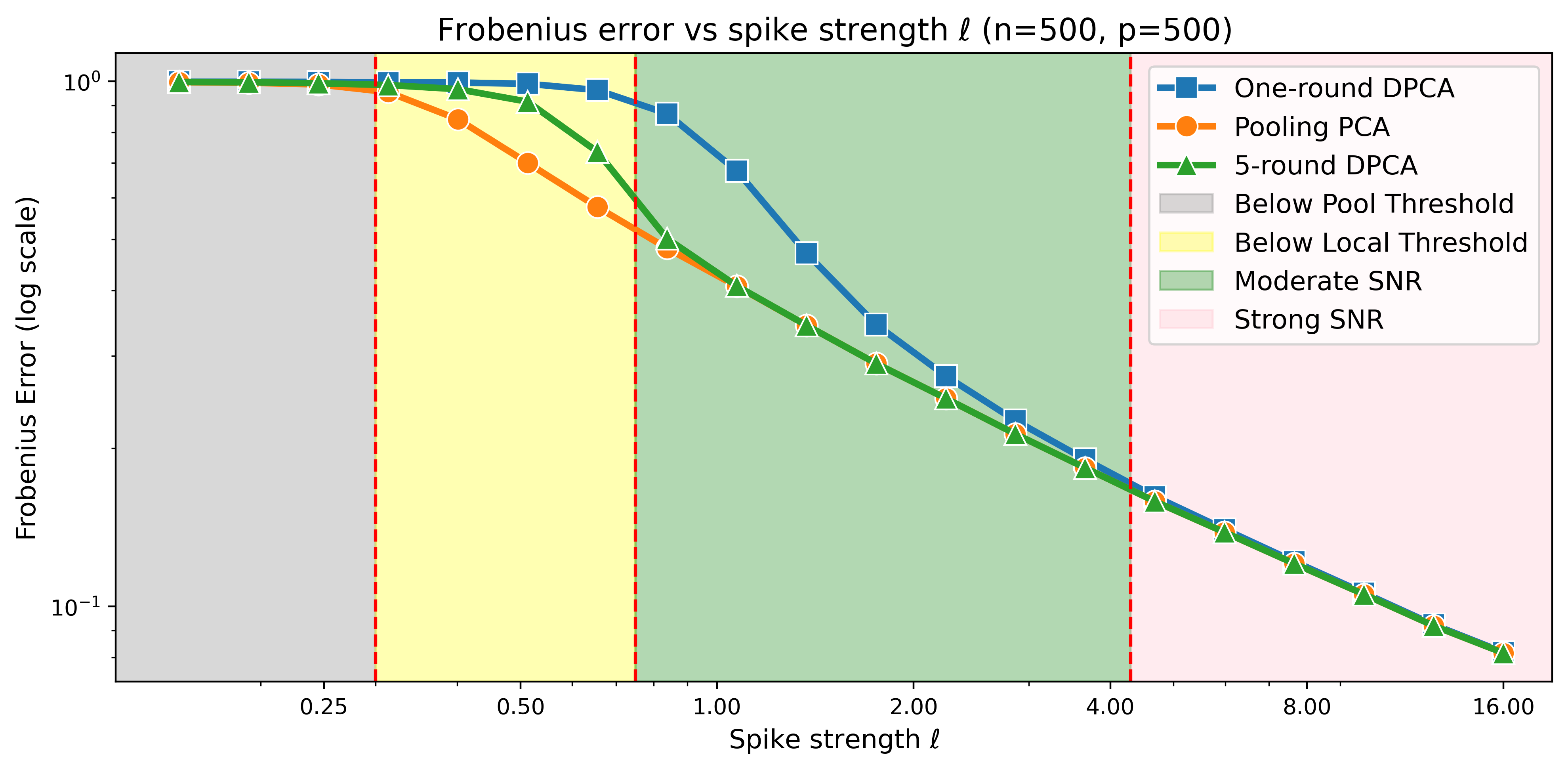}
	\end{minipage}
\caption{Numerical comparisons of different methods under various signal-to-noise regimes of the Gaussian spiked model with $n=p=500$, $r=1$ and $K=10$, averaged based on $100$ replications.\label{fig-phase}}
\end{figure}

In Figure \ref{fig-phase}, we numerically compare the performance of different methods in different signal-to-noise regimes of the Gaussian spiked model. We generate data according to $n=p=500$, $r=1$ and $K=10$. As the signal strength $\l_r$ grows from $0$ to infinity, the numerical results go through the $4$ regimes. They are consistent with the theoretical arguments collected in Table \ref{tab:method_comparison}, demonstrating the theoretical boundaries of different methods and the statistical advantages of the proposed method.

\begin{figure}
	\centering

    \begin{minipage}{0.45\linewidth}
		\centering
	\includegraphics[width=0.85\linewidth]{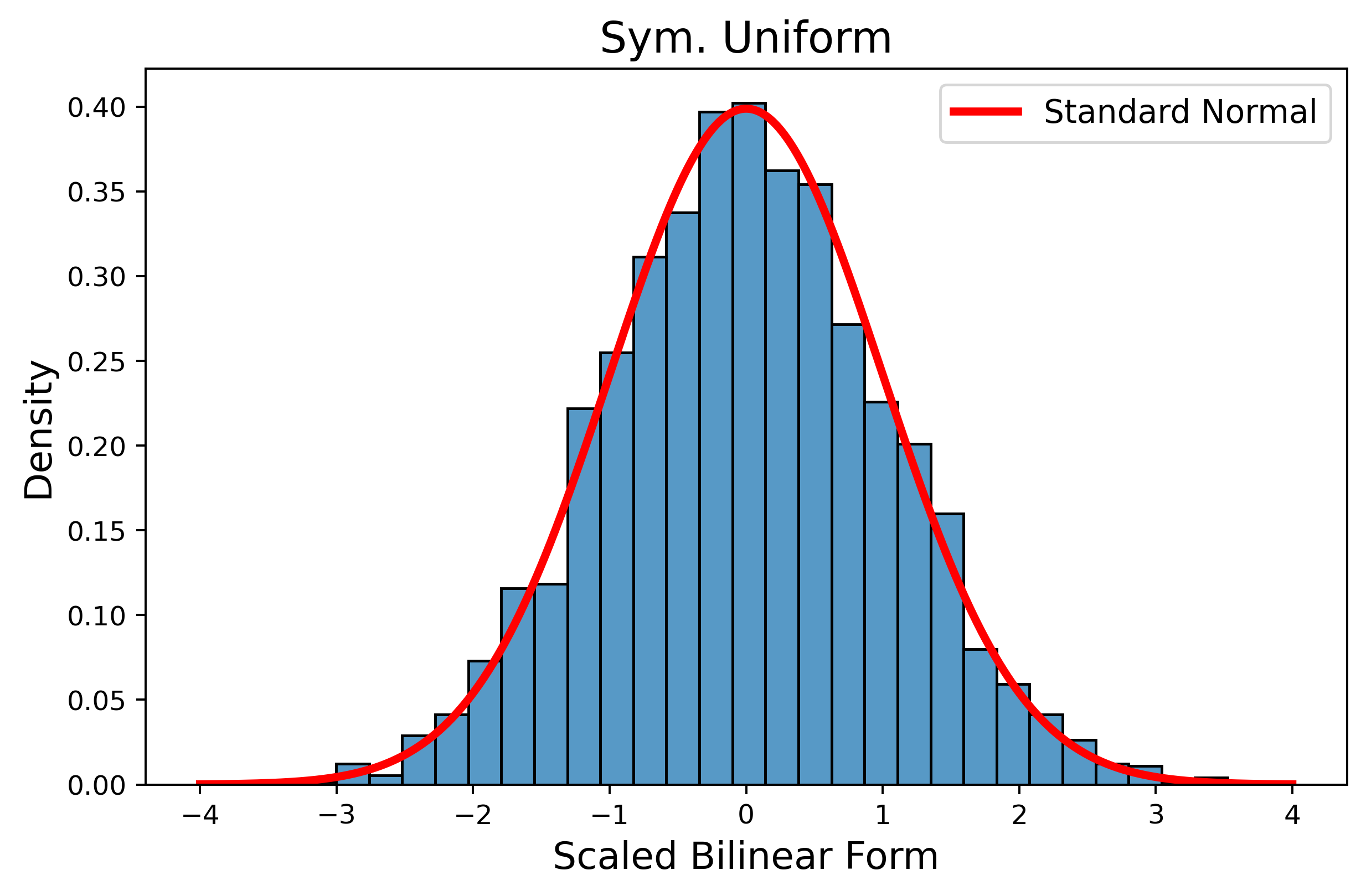}
	\end{minipage}
    \begin{minipage}{0.45\linewidth}
		\centering
	\includegraphics[width=0.85\linewidth]{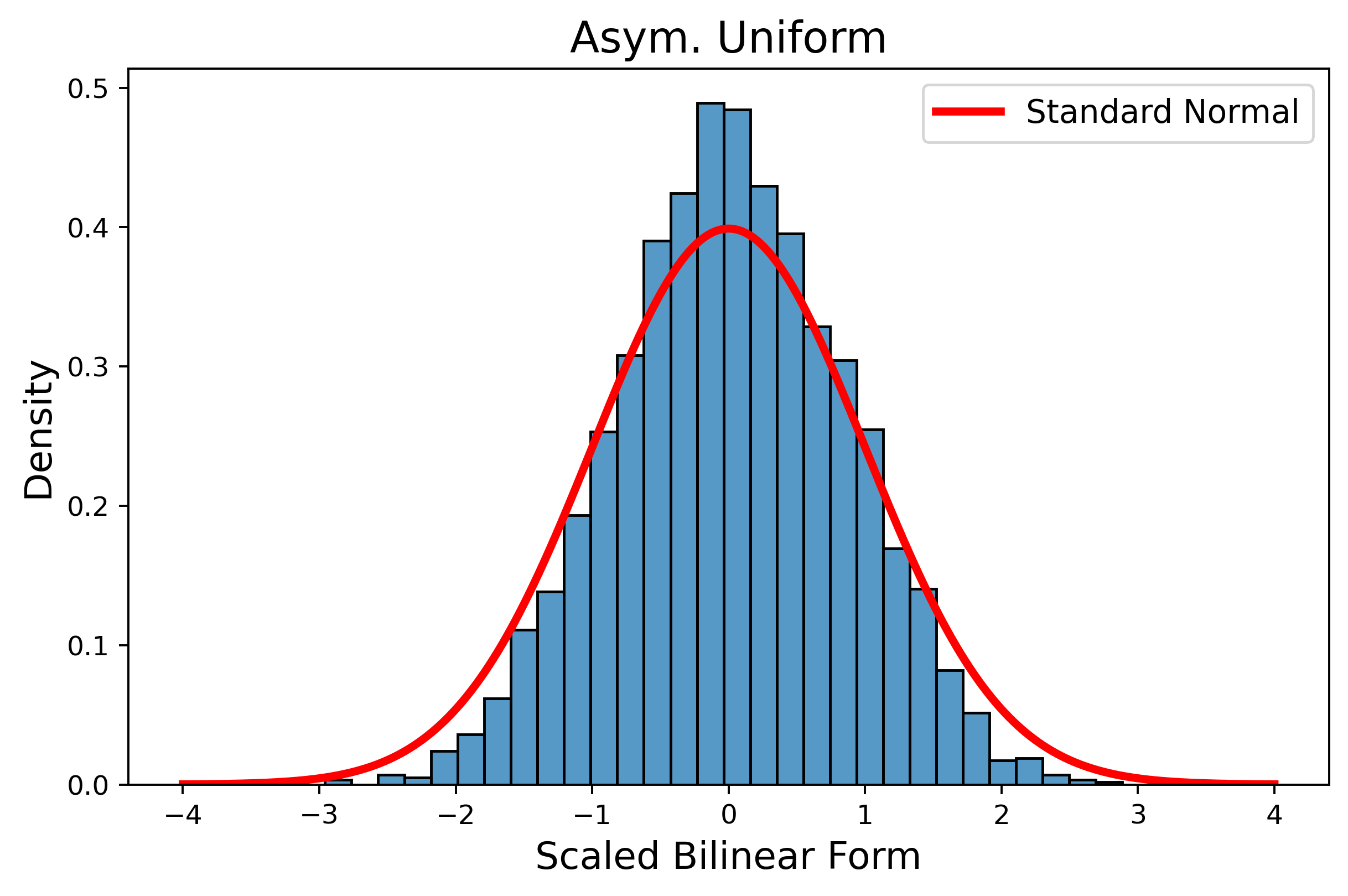}
	\end{minipage}
        \begin{minipage}{0.45\linewidth}
		\centering
	\includegraphics[width=0.85\linewidth]{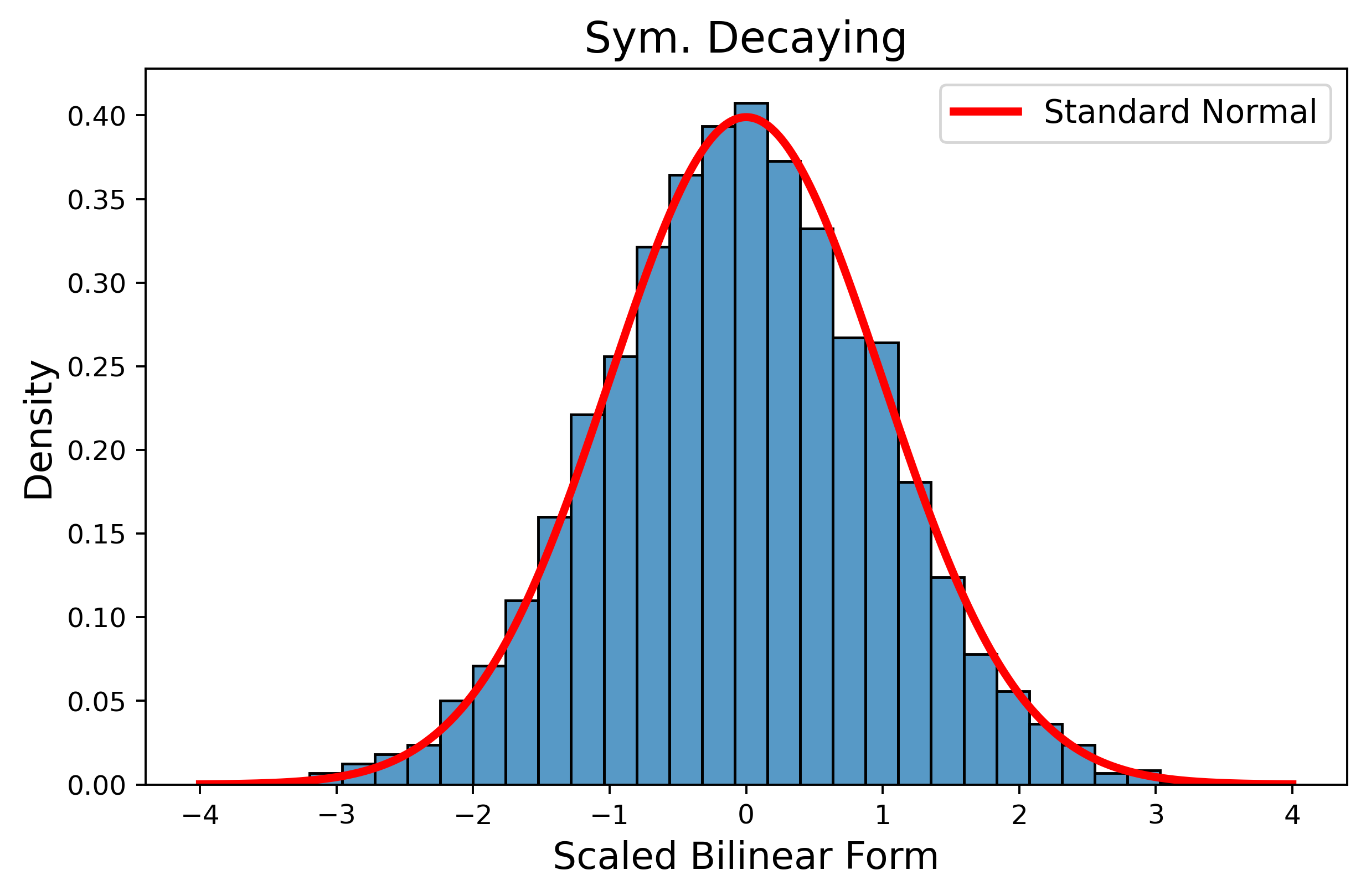}
	\end{minipage}
    \begin{minipage}{0.45\linewidth}
		\centering
	\includegraphics[width=0.85\linewidth]{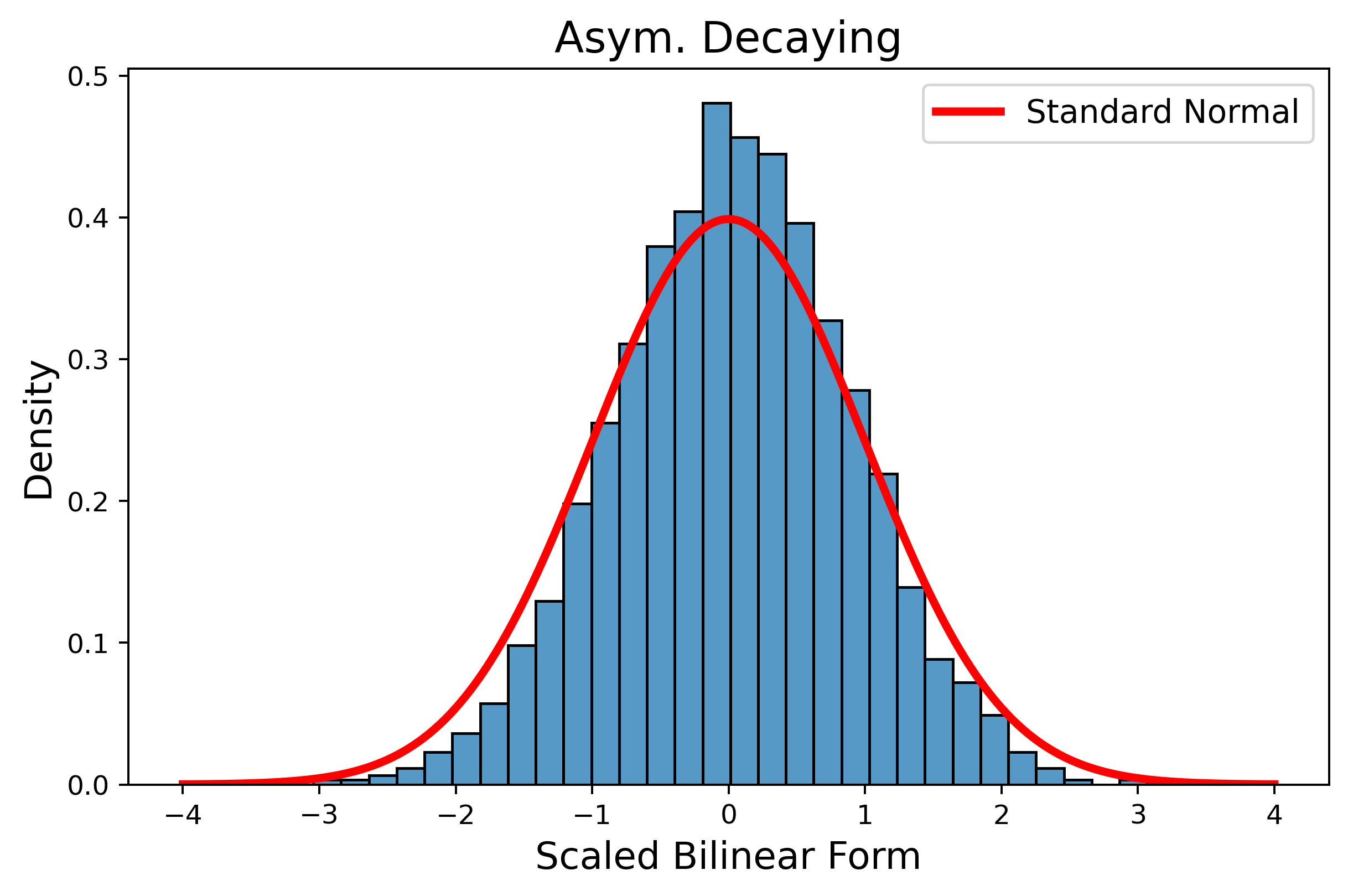}
	\end{minipage}
\caption{Asymptotic normality of $\sqrt{Kn}\langle u, [\hat{U}^{(3)}(\hat{U}^{(3)})^{\top}-UU^{\top}]v\rangle/\hat{\sigma}_{(u,v)}^{(3)}$, for $u$, $v$ generated uniformly from $\in S^{p-1}$ under different scenarios similar to those of Table \ref{tab:results}, each based on 3000 replications. Note that both $\hat{U}^{(3)}$ and $\hat{\sigma}_{(u,v)}^{(3)}$ can be estimated in a distributed manner with three rounds of communication. \label{Fig-AN} }
\end{figure}

In Figure \ref{Fig-AN}, we numerically verify the asymptotic normality of the bilinear form given in Corollary \ref{coro:an}. The data generation procedure is similar to that of Table \ref{tab:results}. We conduct 3000 replicated experiments for each scenario. We set the skewness parameter $\alpha=2$ for signal dimensions, and $\alpha=1$ for noise dimensions. The population covariance matrix has eigenvalues $(5,4,3)$ for the signal components. We set the dimension $p=100$ and distribute the data across $K=50$ machines, with $n=100$ samples per machine. For $u$, $v$ generated uniformly from $\in S^{p-1}$, we rescale $\sqrt{Kn}\langle u, [\hat{U}^{(3)}(\hat{U}^{(3)})^{\top}-UU^{\top}]v\rangle$ by $\hat{\sigma}_{(u,v)}^{(3)}$ defined in \eqref{eq:hatsigma2}, which is estimated from data in a distributed manner. While asymptotic normality aligns well with theoretical claims under symmetric innovation distributions, the empirical distribution shows a sharper peak around zero if the distribution is skewed, suggesting higher kurtosis in the asymmetrical cases.

\begin{table}
\centering
\caption{Comparison of few-round distributed elliptical PCA estimators across different scenarios under heavy-tailed distributions, in terms of the Frobenius norm, based on $100$ replications. Note that $\hat{U}_{\tau}^{(1)}$ is equivalent to the one-round distributed elliptical PCA estimator proposed by \cite{he2022distributed}.}
\label{tab:kendall_results}
\resizebox{0.85\textwidth}{!}{%
\begin{tabular}{cccccc}
\toprule
$t_3$ & Sym. Uniform & Asym. Uniform & Sym. Decaying & Asym. Decaying \\
\midrule
$\hat{U}_{\tau}^{(1)}$ & 0.0516 $\pm$ 0.0068 & 0.0643 $\pm$ 0.0098 & 0.0531 $\pm$ 0.0073 & 0.0663 $\pm$ 0.0114 \\
$\hat{U}_{\tau}^{(2)}$ & 0.0314 $\pm$ 0.0034 & 0.0349 $\pm$ 0.0042 & 0.0318 $\pm$ 0.0032 & 0.0362 $\pm$ 0.0041 \\
$\hat{U}_{\tau}^{(3)}$ & 0.0312 $\pm$ 0.0034 & 0.0346 $\pm$ 0.0041 & 0.0316 $\pm$ 0.0031 & 0.0356 $\pm$ 0.0040 \\
$\hat{U}_{\tau}^{\cP}$ & 0.0309 $\pm$ 0.0033 & 0.0342 $\pm$ 0.0041 & 0.0312 $\pm$ 0.0030 & 0.0352 $\pm$ 0.0039 \\
$\hat{U}^{\cP}$ & 0.2612 $\pm$ 0.4574 & 0.4642 $\pm$ 0.5683 & 0.2567 $\pm$ 0.4266 & 0.4673 $\pm$ 0.5538 \\
\midrule
$t_5$ & Sym. Uniform & Asym. Uniform & Sym. Decaying & Asym. Decaying \\
\midrule
$\hat{U}_{\tau}^{(1)}$ & 0.0414 $\pm$ 0.0043 & 0.0479 $\pm$ 0.0057 & 0.0426 $\pm$ 0.0044 & 0.0494 $\pm$ 0.0060 \\
$\hat{U}_{\tau}^{(2)}$ & 0.0299 $\pm$ 0.0029 & 0.0318 $\pm$ 0.0031 & 0.0305 $\pm$ 0.0026 & 0.0321 $\pm$ 0.0029 \\
$\hat{U}_{\tau}^{(3)}$ & 0.0299 $\pm$ 0.0029 & 0.0316 $\pm$ 0.0031 & 0.0304 $\pm$ 0.0026 & 0.0318 $\pm$ 0.0029 \\
$\hat{U}_{\tau}^{\cP}$ & 0.0295 $\pm$ 0.0028 & 0.0313 $\pm$ 0.0030 & 0.0301 $\pm$ 0.0026 & 0.0315 $\pm$ 0.0028 \\
$\hat{U}^{\cP}$ & 0.0292 $\pm$ 0.0030 & 0.0298 $\pm$ 0.0032 & 0.0300 $\pm$ 0.0026 & 0.0303 $\pm$ 0.0029 \\
\bottomrule
\end{tabular}
}
\end{table}

In Table \ref{tab:kendall_results}, we first report the performance of distributed elliptical PCA under heavy-tailed distributions. The data generation process is similar to the one corresponding to Table \ref{tab:results}, but we use the package \texttt{sstudentt} to generate $t$ and skew-$t$ distributions, instead of Gaussian and skew-Gaussian there. To be more specific, for symmetric distributions, we generate centered multivariate $t_3$ and $t_5$ distributions, corresponding to $\nu=1$, i.e., no skewness. As for asymmetric distributions, we consider skew-$t_3$ and skew-$t_5$ distributions with skewness parameter $\nu=4$ for signal dimensions, and $\nu=2$ for noise dimensions. Under uniform noise, all noise eigenvalues are set to $1$. For decaying noise, the noise eigenvalues linearly decrease from $1.2$ to $0.8$. The population scatter matrix has eigenvalues $(5,3,2)$ for the signal components. We set the dimension $p=200$ and distribute the data across $K=60$ machines, with $n=200$ samples per machine, and $t$ ranges from $1$ to $3$ for the distributed elliptical PCA estimator $\hat{U}_{\tau}^{(t)}$. Note that $\hat{U}_{\tau}^{(1)}$ is equivalent to the one-round distributed elliptical PCA estimator proposed by \cite{he2022distributed}, $\hat{U}^{\cP}$ is the pooling PCA estimator, and $\hat{U}_{\tau}^{\cP}$ is the pooling elliptical PCA estimator.

\bibliographystyle{apalike}
\bibliography{main}

\begin{thebibliography}{}

\bibitem[Bai et~al., 2018]{bai2018consistency}
Bai, Z., Choi, K.~P., and Fujikoshi, Y. (2018).
\newblock Consistency of aic and bic in estimating the number of significant components in high-dimensional principal component analysis.
\newblock {\em The Annals of Statistics}, 46(3):1050--1076.

\bibitem[Bai and Silverstein, 2010]{bai2010spectral}
Bai, Z. and Silverstein, J.~W. (2010).
\newblock {\em Spectral analysis of large dimensional random matrices}, volume~20.
\newblock Springer, New York.

\bibitem[Baik and Silverstein, 2006]{baik2006eigenvalues}
Baik, J. and Silverstein, J.~W. (2006).
\newblock Eigenvalues of large sample covariance matrices of spiked population models.
\newblock {\em Journal of multivariate analysis}, 97(6):1382--1408.

\bibitem[Bao et~al., 2022]{bao2022statistical}
Bao, Z., Ding, X., Wang, J., and Wang, K. (2022).
\newblock Statistical inference for principal components of spiked covariance matrices.
\newblock {\em The Annals of Statistics}, 50(2):1144--1169.

\bibitem[Bhattacharya and Patrangenaru, 2003]{bhattacharya2003large}
Bhattacharya, R. and Patrangenaru, V. (2003).
\newblock Large sample theory of intrinsic and extrinsic sample means on manifolds.
\newblock {\em The Annals of Statistics}, 31(1):1--29.

\bibitem[Bosq, 2000]{bosq2000stochastic}
Bosq, D. (2000).
\newblock {\em Linear processes in function spaces: theory and applications}, volume 149.
\newblock Springer Science \& Business Media.

\bibitem[Cai et~al., 2020]{cai2020limiting}
Cai, T.~T., Han, X., and Pan, G. (2020).
\newblock Limiting laws for divergent spiked eigenvalues and largest nonspiked eigenvalue of sample covariance matrices.
\newblock {\em The Annals of Statistics}, 48(3):1255--1280.

\bibitem[Charisopoulos et~al., 2021]{charisopoulos2021communication}
Charisopoulos, V., Benson, A.~R., and Damle, A. (2021).
\newblock Communication-efficient distributed eigenspace estimation.
\newblock {\em SIAM Journal on Mathematics of Data Science}, 3(4):1067--1092.

\bibitem[Chen and Zhu, 2024]{chen2024distributed}
Chen, C. and Zhu, L. (2024).
\newblock Distributed estimation and gap-free analysis of canonical correlations.
\newblock {\em arXiv preprint arXiv:2412.17792}.

\bibitem[Chen and Fan, 2023]{chen2021statistical}
Chen, E.~Y. and Fan, J. (2023).
\newblock Statistical inference for high-dimensional matrix-variate factor models.
\newblock {\em Journal of the American Statistical Association}, 118(542):1038--1055.

\bibitem[Chen et~al., 2022]{chen2022distributed}
Chen, X., Lee, J.~D., Li, H., and Yang, Y. (2022).
\newblock Distributed estimation for principal component analysis: An enlarged eigenspace analysis.
\newblock {\em Journal of the American Statistical Association}, 117(540):1775--1786.

\bibitem[Couillet and Liao, 2022]{couillet2022random}
Couillet, R. and Liao, Z. (2022).
\newblock {\em Random matrix methods for machine learning}.
\newblock Cambridge University Press.

\bibitem[Crone and Crosby, 1995]{crone1995statistical}
Crone, L. and Crosby, D.~S. (1995).
\newblock Statistical applications of a metric on subspaces to satellite meteorology.
\newblock {\em Technometrics}, 37(3):324--328.

\bibitem[Fan et~al., 2023]{fan2023communication}
Fan, J., Guo, Y., and Wang, K. (2023).
\newblock Communication-efficient accurate statistical estimation.
\newblock {\em Journal of the American Statistical Association}, 118(542):1000--1010.

\bibitem[Fan et~al., 2013]{fan2013large}
Fan, J., Liao, Y., and Mincheva, M. (2013).
\newblock Large covariance estimation by thresholding principal orthogonal complements.
\newblock {\em Journal of the Royal Statistical Society: Series B}, 75(4):603--680.

\bibitem[Fan et~al., 2018]{Fan2018LARGE}
Fan, J., Liu, H., and Wang, W. (2018).
\newblock Large covariance estimation through elliptical factor models.
\newblock {\em The Annals of Statistics}, 46(4):1383--1414.

\bibitem[Fan et~al., 2019]{fan2019distributed}
Fan, J., Wang, D., Wang, K., and Zhu, Z. (2019).
\newblock Distributed estimation of principal eigenspaces.
\newblock {\em Annals of statistics}, 47(6):3009--3031.

\bibitem[Ford, 2014]{ford2014numerical}
Ford, W. (2014).
\newblock {\em Numerical linear algebra with applications: Using MATLAB}.
\newblock Academic Press.

\bibitem[Garber et~al., 2017]{garber2017communication}
Garber, D., Shamir, O., and Srebro, N. (2017).
\newblock Communication-efficient algorithms for distributed stochastic principal component analysis.
\newblock In {\em International Conference on Machine Learning}, pages 1203--1212. PMLR.

\bibitem[Golub and Van~Loan, 2013]{golub2013matrix}
Golub, G.~H. and Van~Loan, C.~F. (2013).
\newblock {\em Matrix computations}.
\newblock JHU press.

\bibitem[Grinsztajn et~al., 2022]{grinsztajn2022tree}
Grinsztajn, L., Oyallon, E., and Varoquaux, G. (2022).
\newblock Why do tree-based models still outperform deep learning on typical tabular data?
\newblock {\em Advances in neural information processing systems}, 35:507--520.

\bibitem[Han and Liu, 2018]{han2018eca}
Han, F. and Liu, H. (2018).
\newblock Eca: High-dimensional elliptical component analysis in non-gaussian distributions.
\newblock {\em Journal of the American Statistical Association}, 113(521):252--268.

\bibitem[He et~al., 2022]{he2022large}
He, Y., Kong, X., Yu, L., and Zhang, X. (2022).
\newblock Large-dimensional factor analysis without moment constraints.
\newblock {\em Journal of Business \& Economic Statistics}, 40(1):302--312.

\bibitem[He et~al., 2025]{he2022distributed}
He, Y., Liu, Z., and Wang, Y. (2025).
\newblock Distributed learning for principal eigenspaces without moment constraints.
\newblock {\em Journal of Computational and Graphical Statistics}, 34(1):318--329.

\bibitem[Huang et~al., 2021]{huang2021communication}
Huang, Z., Lin, X., Zhang, W., and Zhang, Y. (2021).
\newblock Communication-efficient distributed covariance sketch, with application to distributed pca.
\newblock {\em Journal of Machine Learning Research}, 22(80):1--38.

\bibitem[Jolliffe et~al., 2003]{jolliffe2003modified}
Jolliffe, I.~T., Trendafilov, N.~T., and Uddin, M. (2003).
\newblock A modified principal component technique based on the lasso.
\newblock {\em Journal of computational and Graphical Statistics}, 12(3):531--547.

\bibitem[Jordan et~al., 2019]{jordan2019communication}
Jordan, M.~I., Lee, J.~D., and Yang, Y. (2019).
\newblock Communication-efficient distributed statistical inference.
\newblock {\em Journal of the American Statistical Association}.

\bibitem[Kargupta et~al., 2001]{kargupta2001distributed}
Kargupta, H., Huang, W., Sivakumar, K., and Johnson, E. (2001).
\newblock Distributed clustering using collective principal component analysis.
\newblock {\em Knowledge and Information Systems}, 3:422--448.

\bibitem[Koltchinskii and Lounici, 2016]{koltchinskii2016asymptotics}
Koltchinskii, V. and Lounici, K. (2016).
\newblock Asymptotics and concentration bounds for bilinear forms of spectral projectors of sample covariance.
\newblock {\em Annales de l’Institut Henri Poincar{\'e}-Probabilit{\'e}s et Statistiques}, 52(4):1976--2013.

\bibitem[Koltchinskii and Lounici, 2017]{koltchinskii2017new}
Koltchinskii, V. and Lounici, K. (2017).
\newblock New asymptotic results in principal component analysis.
\newblock {\em Sankhya A}, 79:254--297.

\bibitem[Li et~al., 2024]{li2024tpca}
Li, Z., Qin, K., He, Y., Zhou, W., and Xinsheng, Z. (2024).
\newblock Knowledge transfer across multiple principal component analysis studies.
\newblock {\em Available upon request}.

\bibitem[Liski et~al., 2016]{liski2016combining}
Liski, E., Nordhausen, K., Oja, H., and Ruiz-Gazen, A. (2016).
\newblock Combining linear dimension reduction subspaces.
\newblock In {\em Recent advances in robust statistics: Theory and applications}, pages 131--149. Springer.

\bibitem[Marchenko and Pastur, 1967]{marchenko1967distribution}
Marchenko, V.~A. and Pastur, L.~A. (1967).
\newblock Distribution of eigenvalues for some sets of random matrices.
\newblock {\em Matematicheskii Sbornik}, 114(4):507--536.

\bibitem[Naumov et~al., 2019]{naumov2019bootstrap}
Naumov, A., Spokoiny, V., and Ulyanov, V. (2019).
\newblock Bootstrap confidence sets for spectral projectors of sample covariance.
\newblock {\em Probability Theory and Related Fields}, 174(3):1091--1132.

\bibitem[Paul, 2007]{paul2007asymptotics}
Paul, D. (2007).
\newblock Asymptotics of sample eigenstructure for a large dimensional spiked covariance model.
\newblock {\em Statistica Sinica}, 17(4):1617--1642.

\bibitem[Pearson, 1901]{pearson1901liii}
Pearson, K. (1901).
\newblock Liii. on lines and planes of closest fit to systems of points in space.
\newblock {\em The London, Edinburgh, and Dublin philosophical magazine and journal of science}, 2(11):559--572.

\bibitem[Qu et~al., 2002]{qu2002principal}
Qu, Y., Ostrouchov, G., Samatova, N., and Geist, A. (2002).
\newblock Principal component analysis for dimension reduction in massive distributed data sets.
\newblock {\em Proceedings of IEEE International Conference on Data Mining (ICDM)}, 1318(1784--1788).

\bibitem[Schizas and Aduroja, 2015]{schizas2015distributed}
Schizas, I.~D. and Aduroja, A. (2015).
\newblock A distributed framework for dimensionality reduction and denoising.
\newblock {\em IEEE Transactions on Signal Processing}, 63(23):6379--6394.

\bibitem[Shamir et~al., 2014]{shamir2014communication}
Shamir, O., Srebro, N., and Zhang, T. (2014).
\newblock Communication-efficient distributed optimization using an approximate newton-type method.
\newblock In {\em International conference on machine learning}, pages 1000--1008. PMLR.

\bibitem[Silin and Fan, 2020]{silin2020hypothesis}
Silin, I. and Fan, J. (2020).
\newblock Hypothesis testing for eigenspaces of covariance matrix.
\newblock {\em arXiv preprint arXiv:2002.09810}.

\bibitem[Tropp, 2012]{tropp2012user}
Tropp, J.~A. (2012).
\newblock User-friendly tail bounds for sums of random matrices.
\newblock {\em Foundations of computational mathematics}, 12(4):389--434.

\bibitem[Tropp et~al., 2015]{tropp2015introduction}
Tropp, J.~A. et~al. (2015).
\newblock An introduction to matrix concentration inequalities.
\newblock {\em Foundations and Trends in Machine Learning}, 8(1-2):1--230.

\bibitem[Vershynin, 2018]{vershynin2018high}
Vershynin, R. (2018).
\newblock {\em High-dimensional probability: An introduction with applications in data science}, volume~47.
\newblock Cambridge university press.

\bibitem[Wang and Fan, 2017]{wang2017asymptotics}
Wang, W. and Fan, J. (2017).
\newblock Asymptotics of empirical eigenstructure for high dimensional spiked covariance.
\newblock {\em Annals of statistics}, 45(3):1342--1374.

\bibitem[Wu et~al., 2018]{wu2018review}
Wu, S.~X., Wai, H.-T., Li, L., and Scaglione, A. (2018).
\newblock A review of distributed algorithms for principal component analysis.
\newblock {\em Proceedings of the IEEE}, 106(8):1321--1340.

\bibitem[Xia et~al., 2021]{xia2021statistically}
Xia, D., Yuan, M., and Zhang, C.-H. (2021).
\newblock Statistically optimal and computationally efficient low rank tensor completion from noisy entries.
\newblock {\em The Annals of Statistics}, 49(1):76--99.

\bibitem[Xu et~al., 2022]{xu2022distributed}
Xu, K., Zhu, L., and Fan, J. (2022).
\newblock Distributed sufficient dimension reduction for heterogeneous massive data.
\newblock {\em Statistica Sinica}, 32:2455--2476.

\bibitem[Yu et~al., 2015]{yu2015useful}
Yu, Y., Wang, T., and Samworth, R.~J. (2015).
\newblock A useful variant of the davis--kahan theorem for statisticians.
\newblock {\em Biometrika}, 102(2):315--323.

\end{thebibliography}

\end{document}